\documentclass[11pt,a4paper]{article}
\usepackage{jheppub}
\usepackage{mathrsfs}
\usepackage{amssymb}
\usepackage{amsmath}
\usepackage{graphicx}
\usepackage{amsfonts,amsbsy}
\usepackage{nicefrac}
\usepackage{xcolor}

\definecolor{lcolor}{rgb}{0.5,0,0}
\definecolor{citcolor}{rgb}{0,0.3,0.0}
\definecolor{ao(english)}{rgb}{0.0, 0.5, 0.0}
\usepackage{comment}

\newcommand{\mbf}{\mathbf}

\newcommand{\tcent}{\bar{t}}			
\newcommand{\mrm}{\mathrm}

\newcommand{\HTL}{\mrm{HTL}}

\newcommand{\qs}{Q_\mathrm{s}}
\newcommand{\pInit}{p_0}
\newcommand{\Q}{Q}
\newcommand{\wplas}{\omega_{\mrm{pl}}}
\newcommand{\masy}{m_{\mrm{HTL}}}

\newcommand{\SU}{\mrm{SU}}
\newcommand{\su}{\mrm{su}}

\newcommand{\fE}{f_{\mrm{E}}}
\newcommand{\fpi}{f_{\pi}}

\newcommand{\dpol}{d_{\mrm{pol}}}
\newcommand{\dA}{d_{\mrm{A}}}

\newcommand{\fThrD}{f_{\mrm{3D}}}

\newcommand{\pOvMD}{{\tilde{p}}}

\newcommand{\fig}{Fig.~}
\newcommand{\figs}{Figs.~}
\newcommand{\eq}{Eq.~}
\newcommand{\se}{Sec.~}
\newcommand{\eqs}{Eqs.~}
\newcommand{\re}{Ref.~}
\newcommand{\res}{Refs.~}
\newcommand{\app}{Appendix~}
\newcommand{\nr}[1]{(\ref{#1})}
\newcommand{\nc}{N_{\mathrm{c}}}

\newcommand{\ud}{\mathrm{d}}

\newcommand{\dtmax}{\Delta t_{\text{max}}}

\newcommand{\pToFigs}{.}

\newcommand{\tDTwMi}{2D}
\newcommand{\tDThrMi}{2D+sc}
\newcommand{\tThrD}{3D}

\begin{document}

\title{Broad excitations in a 2+1D overoccupied gluon plasma}

\author[a]{K.~Boguslavski} 
\affiliation[a]{Institute for Theoretical Physics, Technische Universit\"{a}t Wien, 1040 Vienna,
Austria}

\author[b]{A.~Kurkela} 
\affiliation[b]{Faculty of Science and Technology, University of Stavanger, 4036 Stavanger, Norway}

\author[c,d]{T.~Lappi} 
\affiliation[c]{Department of Physics, University of Jyv\"{a}skyl\"{a}, P.O.~Box 35, 40014 University of Jyv\"{a}skyl\"{a}, Finland}
\affiliation[d]{Helsinki Institute of Physics, P.O.~Box 64, 00014 University of Helsinki, Finland}

\author[e,f]{J.~Peuron} 
\affiliation[e]{Dept. of Astronomy and Theoretical Physics, S\"{o}lvegatan 14A, S-223 62 Lund, Sweden}
\affiliation[f]{European Centre for Theoretical Studies in Nuclear Physics and Related Areas (ECT*) \\ and Fondazione Bruno Kessler, Strada delle Tabarelle 286, I-38123 Villazzano (TN), Italy}

\emailAdd{kirill.boguslavski@tuwien.ac.at}
\emailAdd{aleksi.kurkela@uis.no}
\emailAdd{tuomas.v.v.lappi@jyu.fi}
\emailAdd{jarkko.peuron@thep.lu.se}

\preprint{LU-TP 21-01}

\abstract{
Motivated by the initial stages of high-energy heavy-ion collisions,
we study excitations of far-from-equilibrium 2+1 dimensional gauge theories using classical-statistical lattice simulations. 
We evolve field perturbations over a strongly overoccupied background undergoing self-similar evolution. While in 3+1D the excitations are described by hard-thermal loop theory, their structure in 2+1D is nontrivial and nonperturbative. 
These nonperturbative interactions lead to broad excitation peaks in spectral and statistical correlation functions. 
Their width is comparable to the frequency of soft excitations, demonstrating the absence of soft quasiparticles in these theories. 
Our results also suggest that excitations at higher momenta are sufficiently long-lived, such that an effective kinetic theory description for 2+1 dimensional Glasma-like systems may exist, but its collision kernel must be nonperturbatively determined. 
}

\arxivnumber{2101.02715}

\maketitle
\flushbottom


\section{Introduction}
\label{sec_introduction}

In the weak coupling framework, the initial stage of a relativistic heavy-ion collision in the high-energy limit is dominated by nonperturbatively strong boost invariant gluon fields \cite{Gelis:2010nm}. At the characteristic momentum scale $\qs$, these strong fields have nonperturbatively large occupation numbers $ f \sim 1/g^2$ and are thus effectively classical \cite{Lappi:2006fp,Gelis:2015gza}.  The success of a hydrodynamic description of the later stages of the evolution indicates that the matter evolves towards sufficient isotropy quickly. 
One of the major open questions in the  field is to understand in a controlled theoretical framework the evolution from a classical field-dominated system with approximately boost-invariant fields at $\tau \lesssim 1/\qs$ to a state close to local thermal equilibrium~\cite{Gelis:2015gza}.

Several studies in the past have addressed the evolution of the initial boost-invariant field during the timescale $\tau \sim 1/\qs$, which can be understood in terms of 2+1-dimensional lattice simulations~\cite{Krasnitz:1998ns,Krasnitz:2001qu,Lappi:2003bi,Krasnitz:2003jw} or analytical calculations~\cite{Lappi:2006fp,Lappi:2006hq,Fries:2006pv,Chen:2015wia}. These approaches
have been combined with hydrodynamic evolution in the later stages, producing a phenomenologically successful descriptions of heavy-ion collisions~\cite{Schenke:2012wb,Schenke:2012hg,Mantysaari:2017cni}. However, these calculations have not been based on a detailed microscopical description of the evolution from boost invariant classical fields to hydrodynamics.

In practice, the initial boost invariance is broken by different mechanisms. At finite collision energy it is broken by the finite thickness of the nucleus \cite{Gelfand:2016yho,Ipp:2017lho,Ipp:2018hai}. Another source of violation are rapidity dependent unstable quantum fluctuations \cite{Romatschke:2003ms,Rebhan:2004ur,Romatschke:2005pm,Epelbaum:2013waa,Gelis:2013rba}. 
After the boost invariance has been broken and the system has become more dilute with an occupation number $1 \ll f \ll 1/g^2$, the classical fields can also be described using kinetic theory \cite{Mueller:2002gd,Jeon:2004dh,Berges:2004yj,Kurkela:2012hp}. This stage has been studied extensively using real time lattice simulations \cite{Berges:2007re,Kurkela:2012hp,Berges:2012ev,Schlichting:2012es,Berges:2013fga,Berges:2013eia,Berges:2013lsa}. More recent studies, which match kinetic theory to hydrodynamics \cite{Kurkela:2015qoa,Kurkela:2018wud,Kurkela:2018vqr}, seem to favor a bottom-up type isotropization scenario~\cite{Baier:2000sb}. However, it is unclear if the kinetic theory approach can be used all the way back to the purely 2+1-dimensional (2+1D) initial stage.
This is one of the questions we aim to address in this paper, which is a continuation of our earlier study of equal-time correlation functions in the same systems in Ref.~\cite{Boguslavski:2019fsb}.

Formulating a kinetic theory in 2+1D faces certain problematic issues \cite{Boguslavski:2019fsb}. The kinetic theory relies on a separation of modes, into soft modes described by classical field theory and hard momentum modes described as particles that undergo scattering. In order to find leading order expressions of the medium-modified scattering cross-sections of the hard particles, the dynamics of the soft sector need to be solved. In three dimensions the dominant medium modification to the soft fields arises from the interaction with the hard modes, and this interaction can be analytically solved giving rise to the Hard-Loop (HTL) theory \cite{Braaten:1989mz,Blaizot:2001nr} and the Debye mass scale $m_D$. 
However, in two spatial dimensions, the lower dimensionality puts more weight on infrared modes in momentum integrals, and consequently, the soft and hard scales contribute equally to the Debye screening mass. As a consequence of their self-interactions, already modes at the soft scale $m_D$ become nonperturbative, bearing resemblance to the magnetic scale in three dimensions. This means that it is not obvious whether the soft modes can be described using HTL theory. This makes it difficult to find leading order accurate matrix elements for kinetic theory in 2+1D analytically.

Nevertheless, our recent studies in 2+1D classical lattice gauge theory \cite{Boguslavski:2019fsb} indicated that 
not only isotropic 3+1D but
also 2+1D systems exhibit a self-similar attractor behavior, for which we extracted the scaling exponents. We also found that the scaling exponents can be understood using a simple kinetic theory analysis. This suggests that 2+1D systems involve quasiparticle excitations, at least at high momenta.

Inspired by these results, our main motivation in this paper is to study the excitation spectrum of a gluonic plasma in extreme anisotropy, such that it is effectively in 2+1 dimensions. For this reason, we work in a theoretically clean environment, which is not complicated by additional phenomenological ingredients, and we will study theories with and without a scalar field in the adjoint representation. Especially the former case resembles the boost invariant initial state of heavy-ion collisions at high energy, as the scalar field arises by dimensional reduction from 3+1 dimensional pure gauge theory. In the future, it will be interesting to study excitation spectra of expanding systems and the influence of different initial conditions that are used to model heavy-ion collisions, while here we are looking at universal properties of the spectrum that are independent of the details of the initial conditions.

In practice, the excitation spectrum is studied using the linear response framework built in \cite{Kurkela:2016mhu} and first applied in \cite{Boguslavski:2018beu} to an isotropic 3+1D gluonic system. A similar classical-statistical linear response framework has been recently used in scalar theories at self-similar attractors \cite{PineiroOrioli:2018hst,Boguslavski:2019ecc} and extends classical-statistical simulations in thermal equilibrium that use a fluctuation-dissipation relation explicitly \cite{Aarts:2001yx,Schlichting:2019tbr}. 
Our work is a natural extension of our previous study of the excitation spectrum of isotropic 3+1D gluodynamics at a classical self-similar attractor \cite{Boguslavski:2018beu}. There we have observed narrow quasiparticle excitation peaks in the spectrum for all momenta and a generalized fluctuation-dissipation relation between spectral and statistical correlation functions. 

While we also find a generalized fluctuation-dissipation relation, our main result is
 that the gluonic correlation functions in 2+1D systems exhibit a broad excitation peak of non-Lorentzian shape, in contrast to 3+1D. In fact, we show that the peak width is of the order of and scales with the mass. We separately confirm in Appendix~\ref{app_class} that this happens also in classical thermal equilibrium. 
This indicates that gluonic excitations for momenta below the mass are too short-lived to form quasiparticles. Moreover, expressions obtained from HTL theory mostly do not provide a good description of our simulation results. This is in line with the breakdown of the HTL theory at the soft scale. The interaction between soft excitations is genuinely nonperturbative in 2+1D, and the dynamical description has to take this into account. Our results also suggest that gluonic and scalar excitations at higher momenta are sufficiently long-lived, such that we may have an effective kinetic description also in the strongly anisotropic systems which arise in the early stages of ultrarelativistic heavy ion collisions, albeit with nonperturbatively detemined collision kernels.

We start in \se\ref{sec_theory} by introducing the considered theories and relevant correlation functions, after which we describe our numerical method and introduce the main HTL expressions. Our numerical results for non-equilibrium 2+1D systems are shown in \se\ref{sec:results}.
We conclude in \se \ref{sec:conclu}. The analytical forms for the HTL functions in 2+1D are derived in Appendix~\ref{app_HTL2D}, and a numerical study in classical thermal equilibrium is discussed in Appendix~\ref{app_class}.


\section{Theoretical background}
\label{sec_theory}
\subsection{3+1D, 2+1D and Glasma-like 2+1D theories}

We consider non-Abelian $\SU(N_c)$ gauge theories with $N_c = 2$ in $d$ spatial dimensions. Their classical action reads 
\begin{align}
 \label{eq_class_action}
 S_{\mrm YM}[A] = -\frac{1}{4}\,\int d^{d+1}x\;F_a^{\mu\nu} F^a_{\mu\nu},
\end{align}
with field strength tensor $F^a_{\mu\nu} = \partial_\mu A^a_\nu - \partial_\nu A^a_\mu + g\, f^{abc}\, A^b_\mu A^c_\nu$, where repeated color indices $a = 1, ..., N_c^2 - 1$ and Lorentz indices $\mu, \nu = 0, ..., d$ imply summation over them. Using the generators $\Gamma^a$ of the $\su(N_c)$ algebra, the gauge field can be written as a fundamental representation matrix $A_\mu = A^a_\mu \Gamma^a$. We consider the following theories:

\begin{itemize}
 
 \item `\tThrD' or `3+1D': in $d = 3$ spatial dimensions, such that
 \begin{align} 
  S_{\mrm YM}[A] = S_{\mrm YM}^{3D}[A]\,.
 \end{align}
 
 \item `\tDTwMi' or `2+1D': in $d = 2$ spatial dimensions, such that
 \begin{align} 
  S_{\mrm YM}[A] = S_{\mrm YM}^{2D}[A]\,.
 \end{align}

 \item `\tDThrMi' or `Glasma-like 2+1D': originally in $d = 3$ spatial dimensions where no field depends on the coordinate $x^3$. This results in a 2+1D theory with the classical action
 \begin{align}
  S_{\mrm YM}[A] = \Q L_3 \left(S_{\mrm YM}^{2D}[A] + S_{\phi}^{2D}[\phi]\right)
 \end{align}
 with an adjoint scalar field $\phi^a \equiv A_3^a$ having the action
\begin{align}
 S_{\phi}^{2D}[\phi] = -\frac{1}{2}\,\int d^{2+1}x\;(D^{ab}_{\mu}\phi^b) (D_{ac}^{\mu}\phi^c),
\end{align}
with summation over $\mu =0, 1,2$ and with the covariant derivative $D^{ab}_\mu = \delta^{ab}\partial_\mu - g f^{abc} A_\mu^c$. The length in the $x^3$ direction $L_3$ drops out of the classical dynamics. We have also factored out a constant momentum scale $\Q$ and rescaled the gauge coupling and all fields as $g\,\Q^{1/2} \rightarrow g$, $A\,\Q^{-1/2} \rightarrow A$, $\dots$, such that the momentum dimensions of the action, the coupling constant and the fields match the ones for 2+1D. This results in $[S_{\mrm YM}] = [S_{\mrm YM}^{2D}] = [S_{\phi}^{2D}] = 0$, $[g] = 1/2$ and $[A] = [\phi] = 1/2$ for both 2D theories. The momentum scale $\Q$ used in the rescaling is in principle arbitary, but we will use a value related to the conserved energy density, as discussed in more detail below. 

We will refer to this theory as Glasma-like 2+1D, since the Glasma occurring at the initial stages at ultrarelativistic heavy ion collisions is also invariant in one spatial dimension (rapidity) \cite{Lappi:2003bi,Lappi:2006fp,Lappi:2006hq}. 
Unlike the expanding Glasma usually employed in models of the initial collision dynamics, the 2+1D theory we consider here is defined in a nonexpanding coordinate system, and our initial condition corresponds to a positive (but small) longitudinal pressure $P_L > 0$. Also in the expanding Glasma the expansion becomes less important at later times, as reflected e.g.\ in the fact that $P_L$ becomes positive at $\tau \gtrsim 1/\qs$. Thus our Glasma-like system could be interpreted as a nonexpanding model of this later Glasma state.
\end{itemize}

\subsection{Spectral and statistical correlation functions}
\label{sec:theory_corrs}

Here we will give a brief overview of the spectral and statistical correlation functions measured in this work. For a more comprehensive introduction and description of the methods used we refer the reader to \cite{Boguslavski:2018beu}.

The statistical correlation function is defined as the anticommutator of two field operators
\begin{align}
 \label{eq_stat_fct}
 \langle AA \rangle_{jk} (x, x') &= \frac{1}{2\dA}\sum_a\left\langle \left\{ \hat{A}_j^a(x), \hat{A}_k^a(x') \right\} \right\rangle \nonumber \\
 \langle EE \rangle^{jk} (x, x') &= \frac{1}{2\dA}\sum_a\left\langle \left\{ \hat{E}^j_a(x), \hat{E}^k_a(x') \right\} \right\rangle,
\end{align}
with spatial components $j,k = 1, \dots, d$, the dimension of the adjoint representation $\dA = \nc^2-1$ and spacetime coordinates $x \equiv (t,\mbf x)$. Because of $E^j(x) = \partial_t A_j(x)$, the correlators $\langle EE \rangle$ and $\langle AA \rangle$ are related via time derivatives.
In analogy, the statistical correlation function for scalar components of the Glasma-like theory are defined as $\langle \phi\phi \rangle (x, x') = \frac{1}{2\dA}\sum_a\left\langle \left\{ \hat{\phi}^a(x), \hat{\phi}^a(x') \right\} \right\rangle$ and $\langle \pi\pi \rangle (x, x') = \frac{1}{2\dA}\sum_a\left\langle \left\{ \hat{\pi}^a(x), \hat{\pi}^a(x') \right\} \right\rangle$. The scalar fields are related to the fields of the original theory via $\phi \equiv A_3$ and $\pi \equiv E^3$.
In the classical limit the statistical correlation functions are easy to measure, since the anticommutator of Heisenberg field operators reduces to a product of two classical fields. 

The spectral function, on the other hand, is defined as the commutator of two field operators
\begin{align}
 \label{eq_spectral_fct}
 \rho_{jk} (x, x') &= \frac{i}{\dA}\sum_a\left\langle \left[ \hat{A}_j^a(x), \hat{A}_k^a(x') \right] \right\rangle  \nonumber \\
 \dot{\rho}_{jk} (x, x') &= \frac{i}{\dA}\sum_a\left\langle \left[ \hat{E}_j^a(x), \hat{A}_k^a(x') \right] \right\rangle, 
\end{align}
and analogously for the scalar components of the Glasma-like theory. We will mostly study its time derivative $\dot{\rho}$, which we will refer to as the ``dotted spectral function''.
In the classical limit the commutator corresponds to the Dirac bracket, which generalizes the Poisson bracket for systems with constraints (such as the Gauss' law). Therefore, a direct measurement of the spectral function is a quite involved task. However, the spectral function is intimately related to the retarded propagator
\begin{align}
 \label{eq_GR_rho_relation}
 G^{R}_{jk}(t,t',p) = \theta(t - t')\, \rho_{jk}(t,t',p). 
\end{align}
The latter can be extracted numerically by using our linear response framework~\cite{Kurkela:2016mhu,Boguslavski:2018beu}. There we introduce a small instantaneous source $j$ for different momentum modes on top of the classical fields $A(x)$, $E(x)$. 
The source generates small response fields $a(x)$, $e(x)$. These follow linearized equations of motion constructed to satisfy the Gauss law constraint exactly. The spectral function $\rho$ and the dotted spectral function $\dot{\rho}$ are computed as correlation functions of $j$ with $a$ and $e$, respectively. 

In \eqref{eq_GR_rho_relation} we implied a spatial Fourier transform with respect to $\mbf x - \mbf x'$. Apart from working in momentum space, it is beneficial to compute the correlation functions in frequency space. For that, a Fourier transform with respect to the relative time $\Delta t = t-t'$ for fixed central time $\tcent = (t+t')/2$ has to be performed, respecting the even parity of $\langle EE \rangle(t,\Delta t,p)$ and $\dot{\rho}(t,\Delta t,p)$ with respect to $\Delta t$. In practice, as discussed in \res\cite{Boguslavski:2018beu,Boguslavski:2020tqz}, we approximate this by a numerically more efficient transform where the lower limit $t$ is kept fixed, and one Fourier transforms with respect to an upper limit $t+\Delta t$:
\begin{align}
 \label{eq:FTrafo}
 \langle EE \rangle(t,\omega,p) = \;&2\int_0^\infty \ud \Delta t \cos(\omega \Delta t)\,\langle E(t+\Delta t/2)E(t-\Delta t/2) \rangle\left(p\right) \nonumber \\
 \approx \;&2\int_0^{\dtmax} \ud \Delta t \, \cos(\omega \Delta t)\,h(\Delta t)\,\langle E(t+\Delta t)E(t) \rangle(p)\,,
\end{align}
and analogously for $\dot{\rho}$. We justify this approximation by the observation that our systems at sufficiently late times depend on $t$ only weakly within a relative time window of the order of the inverse plasmon mass $\sim 1/\wplas$, which is the relevant timescale for $\Delta t$. 
If not stated otherwise, the resulting curves are made smoother by employing standard signal processing techniques, as in \re\cite{Boguslavski:2019ecc}. To this end we have introduced in \eq\nr{eq:FTrafo} a Hann window function
\begin{align}
 h(\Delta t) = \frac{1}{2} \left( 1 + \cos \frac{\pi \Delta t}{\dtmax} \right)
\end{align}
and use zero padding, which implies evaluating \eq\eqref{eq:FTrafo} at more intermediate frequencies than computed using the discrete Fourier transform. As will be demonstrated in \se\ref{sec:peaks}, these methods do not change the forms of the correlation functions considerably, while reducing background ringing that results from a finite time window in the Fourier transform.

In this work, we will not Fourier transform the correlation functions $\langle AA \rangle$ and $\rho$ for 2+1D theories directly, because these functions turn out to oscillate around a non-zero value in the time domain $\Delta t$. 
In stead, we obtain the frequency space spectral function by Fourier-transforming the time derivative $\dot{\rho}(t,\Delta t,p)$, and subsequently dividing by $\omega$. Note that $\rho(t,\Delta t,p)$ and $\rho(t,\omega,p)$ are odd functions in $\Delta t$ and $\omega$ respectively, and are related by a Fourier sine transform in stead of the cosine transform \eqref{eq:FTrafo}. Thus we have explicitly  $\rho(t,\omega{=}0,p)=0$. However, as we will see in Sec.~\ref{sec:smallomega}, the oscillations around a non-zero value at large $\Delta t$ lead to the actual spectral function not approaching this  limit smoothly when $\omega\to 0$.

We will be interested in these correlation functions for transverse and longitudinal polarizations of the gluon field. In 2 spatial dimensions, they can be computed in momentum space by projections on transverse or longitudinal vectors, respectively, as for example in $\langle EE \rangle_T = v_T^j v_T^k \langle EE \rangle_{jk}$ or $\langle EE \rangle_L = v_L^j v_L^k \langle EE \rangle_{jk}$, with $v_T = (p_y,-p_x)^T/p$ and $v_L=(p_x,p_y)^T/p$.

The (single-particle) distribution function is an important quantity that we can use to characterize the dynamical state of the system. We will use the same definitions as in our previous publications \cite{Boguslavski:2018beu,Boguslavski:2019fsb,Boguslavski:2020tqz}
\begin{align}
 \fE(t,p) \,&= \frac{\langle EE \rangle_{T}(t,\Delta t {=} 0,p)}{p} \nonumber \\
 \fpi(t,p) \,&= \frac{\langle \pi\pi \rangle(t,\Delta t {=} 0,p)}{p}\,.
\end{align}

The employed temporal gauge leaves room for  gauge transformations that only depend on the spatial coordinates. Since all the correlation functions discussed in this subsection are not manifestly gauge-invariant observables, we remove the residual gauge freedom for equal time measurements by fixing to Coulomb-type gauge $\left. \partial^j A_j = 0 \right|_{t}$ at the time $t$ of the measurement, as often employed in classical-statistical gauge simulations \cite{Berges:2007re,Kurkela:2012hp,Berges:2012ev,Schlichting:2012es,Berges:2013fga,Berges:2013eia}. In this physical gauge, gauge fields are always transversely polarized while electric fields may include longitudinal terms. For unequal-time correlators we also impose the same condition at the initial time $t$ for the measurement. During the time evolution, the system then gradually shifts away from the gauge condition so that it is not exactly satisfied at the second measurement time $t+\Delta t$.
However, as argued in \cite{Boguslavski:2018beu}, this effect is expected to be small for sufficiently short relative times $\Delta t \ll t$, which we also assumed in order to perform the Fourier transform \eqref{eq:FTrafo} in a finite $\Delta t$ interval.

\subsection{Self-similarity and nonthermal fixed points}
\label{sec:self-sim}
Many highly occupied systems exhibit non-thermal fixed points (also referred to as universal classical attractors). These have been seen in classical non-Abelian gauge theories \cite{Kurkela:2012hp,Berges:2012ev,Schlichting:2012es,Boguslavski:2019fsb,Berges:2019oun}, scalar theories \cite{Berges:2008wm,Orioli:2015dxa,Walz:2017ffj,Berges:2017ldx,Chantesana:2018qsb}, longitudinally expanding systems \cite{Berges:2013eia,Berges:2014bba,Berges:2015ixa} and recently also in ultra-cold atom experiments \cite{Prufer:2018hto,Erne:2018gmz,Glidden:2020qmu}. A characteristic feature of these attractors is that the system forgets the details of its initial conditions and can usually be understood using a very simple set of scaling functions and power laws, which tremendously simplifies the theoretical description. 

In our previous work \cite{Boguslavski:2019fsb}, we established that pure gauge theories also in 2+1D exhibit self-similarity, as defined by 
\begin{align}
 \label{eq_selfsim_1}
 f(t,p) = (Qt)^{\alpha} f_s((Qt)^{\beta} p)\,,
\end{align}
and extracted the universal scaling exponents 
\begin{align}
\label{eq:scaling_exp}
 \beta = -1/5\,, \qquad \alpha = 3\beta\,.
\end{align}
We observed there that the scaling exponents in the 2+1D systems can be understood using relatively simple kinetic theory arguments, even if a kinetic theory description is not expected to work quantitatively. This result was one of the main motivations for the present paper, since it hints that a quasiparticle description of 2+1D theories at high momenta might be possible. 

To understand the physical meaning of \eqs\eqref{eq_selfsim_1} and~\eqref{eq:scaling_exp}, we define the time-dependent hard scale $\Lambda(t)$ as the momentum scale that contributes the most to the perturbative estimate of the energy density $\varepsilon \sim \int \ud^2 p \,p\,f(t, p)$. Then the scaling relation \eqref{eq_selfsim_1} and the scaling exponents imply
\begin{align}
\label{eq:scaling_scales}
 \Lambda(t) \sim \Q (\Q t)^{-\beta}, \quad m_D(t) \sim \Q (\Q t)^{\beta}, \quad \varepsilon = const\,,
\end{align}
where we included the expected scaling of the soft scale represented by the Debye mass from \eq\eqref{eq_DebyeM}. Thus, the hard scale grows and the soft scale decreases with time, while energy density is conserved. 

Throughout this work all the measurements are performed at sufficiently large values of time $t$ to be in the self-similar regime. 

\subsection{Initial conditions}
For the 2+1D systems, we use the same initial condition as in \cite{Boguslavski:2019fsb}. We consider weakly coupled $g^2/\Q \ll 1$ but highly 
occupied $f \gg 1$ systems.   The initial  single-particle distribution function $f(t,\mbf p)$ at the initial time $t = 0$ both for gauge and scalar excitations (i.e., $\fE = \fpi = f$) is given by 
\begin{align}
 \label{eq_2D_IC}
 f(t=0, p) = \frac{\Q}{g^2}\, n_0\, e^{-\frac{p^2}{2\pInit^2}}.
\end{align}
 Here $\Q$ is a gauge invariant  momentum scale, which is more precisely defined by 
 \begin{align}
 \label{eq_Q_def}
 \Q \equiv \sqrt[4]{\frac{C\,g^2 \varepsilon}{\dpol\,(\nc^2-1)}}.
\end{align}
 We consider systems in a fixed size box, and thus the energy density is a conserved quantity. Since on the classical level it is possible to carry out field simulations without making explicit reference to the precise value of the coupling $g$, the quantities we have access to are the energy density scaled with the coupling $g^2 \varepsilon$, which has the momentum dimension $[g^2\varepsilon] = 4$, and $g^2 f$.
Unless stated otherwise, we will use the initial occupation number $n_0 = 0.1,$ for which $\pInit = \Q$ for our chosen definition as detailed below. We have previously shown \cite{Boguslavski:2019fsb} that the form of the initial conditions is unimportant since the systems will approach an attractor solution that only depends on $\Q$ in the sense of \eq\eqref{eq_selfsim_1}. The number of non-longitudinal polarizations in \eqref{eq_2D_IC} is $\dpol$, with $\dpol = 1$ for the 2+1D and $\dpol = 2$ for the Glasma-like 2+1D theories. The constant $C$ is taken as $C = 20 \sqrt{2\pi} \approx 50,$ which we merely choose for convenience such that for $n_0 = 0.1$ one has $\pInit = \Q$. 

In 2+1D the coupling constant $g$ is dimensionful: if one keeps the \emph{dimensionless} combination $g^2/Q$ constant, one observes that \nr{eq_Q_def} leads to the proportionality $\Q \propto \sqrt[3]{\varepsilon},$ which  is natural for a scale derived from a 2-dimensional energy density.
Combining the definition \eqref{eq_Q_def} with a perturbative estimate  $\varepsilon \approx \dpol (\nc^2-1)\,\int \ud^2 p/(2\pi)^2 \,p\,f(t=0, p)$ for the energy density, one obtains
\begin{align}
 \label{eq_Q_def_fGen}
 \Q^3 \,&\approx C \int \frac{\ud^2 p}{(2\pi)^2}\,p\,\dfrac{g^2f(t=0,p)}{\Q} \nonumber \\
 &= 10\, n_0\, \pInit^3\,,
\end{align}
which is independent of $g^2$, $\dpol$ and $\nc$. 

For the 3+1D theory we employ the same isotropic initial condition as in \cite{Boguslavski:2018beu} with the distribution function initially given by
\begin{align}
 f(t=0, p) = \frac{n_0}{g^2}\,\frac{\pInit}{p}\, e^{-\frac{p^2}{2\pInit^2}}.
\end{align}
Here the coupling is dimensionless. In this work we use $n_0=0.2$ for the spatially three-dimensional case and define the characteristic momentum scale as $\Q = \sqrt[4]{5n_0}p_0 \propto \sqrt[4]{g^2 \varepsilon}$.

In our figures, all dimensionful quantities are rescaled by appropriate powers of $\Q$ of the respective theory to make them dimensionless, unless a different rescaling prescription is stated explicitly.

\subsection{HTL expressions}
\label{sec:HTL}

Diagrammatically, the HTL approximation corresponds to an all-order resummation with the kinematic approximation that the external lines are considered to be soft compared to the hard momenta flowing in the internal lines of the diagrams. This corresponds to the non-Abelian generalization of the Vlasov equations, the Wong equations, where the soft modes are considered to be classical fields and the hard modes correspond to classical particles \cite{Blaizot:2001nr}. While we expect the interaction between the soft and hard modes to be described by the HTL theory as is the case in 3+1D, the parametric counting for 2+1D in \cite{Boguslavski:2019fsb} suggests that in addition to the interaction between soft and hard modes, there is also a leading-order interaction between the soft modes among themselves, which is not captured by the HTL approximation. Therefore, we expect deviations from HTL expressions arising from the nonperturbative soft-soft interactions. We compare our numerical results to the HTL calculations to quantify the deviations. 

Here we summarize the main HTL results that are relevant for the comparison with our data. Details are written in  Appendix~\ref{app_HTL2D}. 
The HTL spectral function can be decomposed into a transversely polarized, a longitudinally polarized and, for the Glasma-like theory, a scalar contribution, denoted below by the index $\alpha = T$, $L$, or $\phi$, respectively. 
Each contribution can be further split into a Landau damping part for $|\omega| < p$ and a quasiparticle part 
\begin{align}
\label{eq:rhoHTL_decomposition}
 \rho_{\alpha}^\HTL(\omega,p) = \rho_{\alpha}^{\mrm{Landau}}(\omega,p) + 2\pi Z_{\alpha}(p)\left[ \delta\!\left(\omega - \omega_{\alpha}^\HTL(p)\right) - \delta\!\left(\omega + \omega_{\alpha}^\HTL(p)\right) \right].
\end{align}
The Landau damping expressions in 2+1D read
\begin{align}
\label{eq_Landau_damping_trans}
 \rho_T^{\mrm{Landau}} =&\, \dfrac{2}{m_D^2}\, \dfrac{x\sqrt{1-x^2}\; \theta(1-x^2)}{\left(\left(1-x^2\right)(p/m_D)^2 + x^2\right)^2 + x^2\left(1-x^2\right)} \\
 \label{eq_Landau_damping_long}
 \rho_L^{\mrm{Landau}} =&\, \dfrac{2}{x}\, \dfrac{m_D^2 /\sqrt{1-x^2}\; \theta(1-x^2)}{(p^2 + m_D^2)^2 + x^2 m_D^4/(1-x^2)} \\
 \label{eq_Landau_damping_phi}
 \rho_\phi^{\mrm{Landau}}=&\,0 \,,
\end{align}
with $x = \omega/p$. The Debye mass $m_D$ entering the expressions is determined within HTL at leading order in \eq\eqref{eq_DebyeM} and depends on the distribution function $f(t,p)$. It is connected with the asymptotic mass via $m_D^2 = m_\HTL^2$ in 2+1D theories and $m_D^2 = 2 m_\HTL^2$ for the 3+1D system. In this work, we compute the asymptotic mass like in \cite{Boguslavski:2018beu} in $d$ spatial dimensions as
\begin{align}
 \label{eq_mHTL_selfconst}
 m_\HTL^2 = \dpol N_c \int \frac{\ud^d p}{(2\pi)^d} \frac{g^2f(t,p)}{\sqrt{m_\HTL^2 + p^2}}\,,
\end{align}
which is a self-consistent generalization of \eq\eqref{eq_DebyeM} and reduces the dependence on the definition of the distribution function \cite{Boguslavski:2018beu}.

The dispersion relations in 2+1D are
\begin{align}
\label{eq_HTL_trans}
 \omega_T^\HTL(p) \,&= m_\HTL\,\sqrt{\frac{1 + 2\pOvMD^2 - 2 \pOvMD^4 + \sqrt{1 + 4\pOvMD^2}}{4 - 2\pOvMD^2}} \\ 
 \label{eq_HTL_long}
 \omega_L^\HTL(p) \,&= m_\HTL\,\frac{1+\pOvMD^2}{\sqrt{2+\pOvMD^2}} \\
 \label{eq_HTL_scalar}
 \omega_\phi^\HTL(p) \,&= \sqrt{m_\HTL^2 + p^2}\,,
\end{align}
where we defined $\pOvMD = p / m_\HTL$. The expressions for the quasiparticle residues $Z_\alpha(p)$ are written in Appendix~\ref{app_HTL2D}.

\begin{figure}[t]
	\centering
	\includegraphics[scale=0.06]{\pToFigs/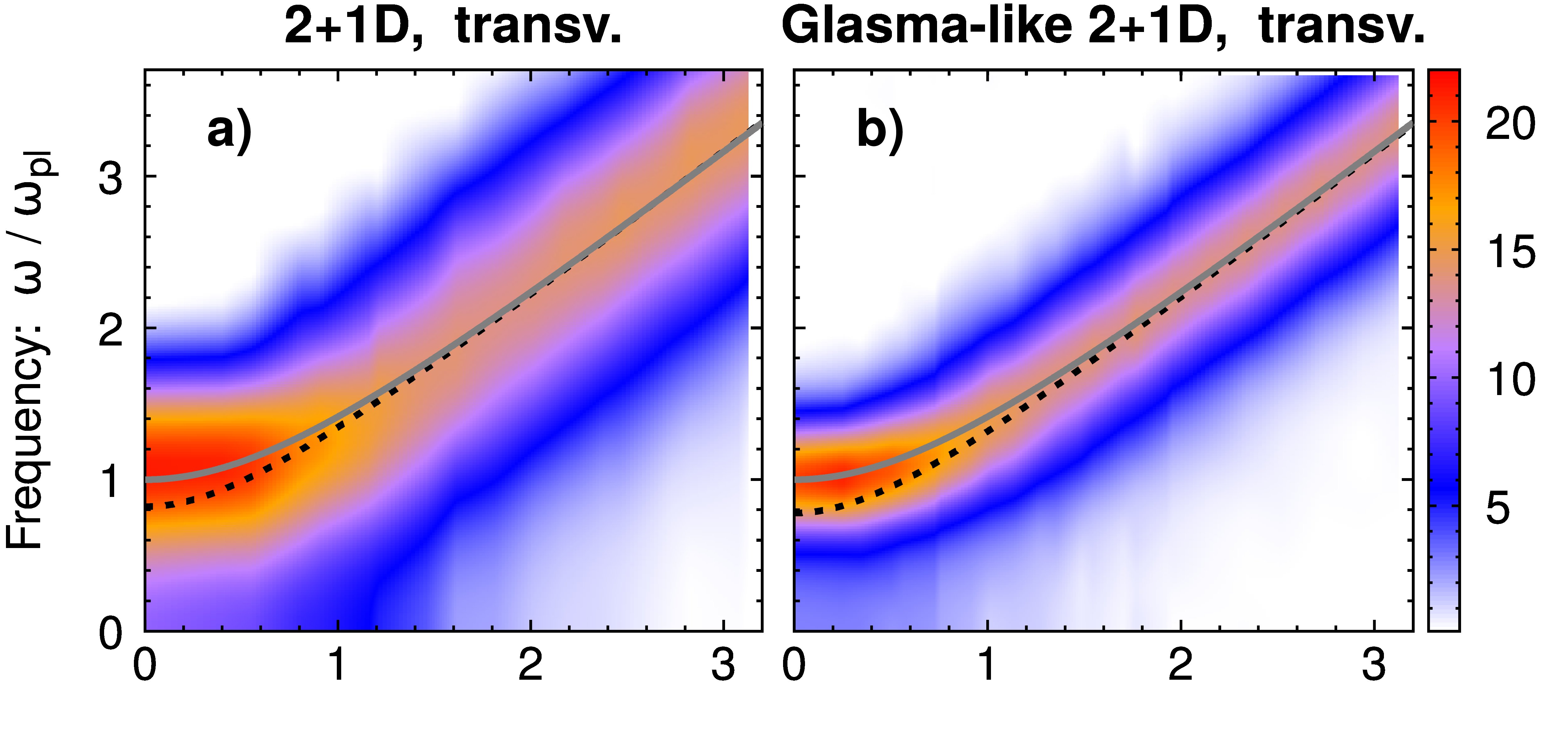}
	\includegraphics[scale=0.06]{\pToFigs/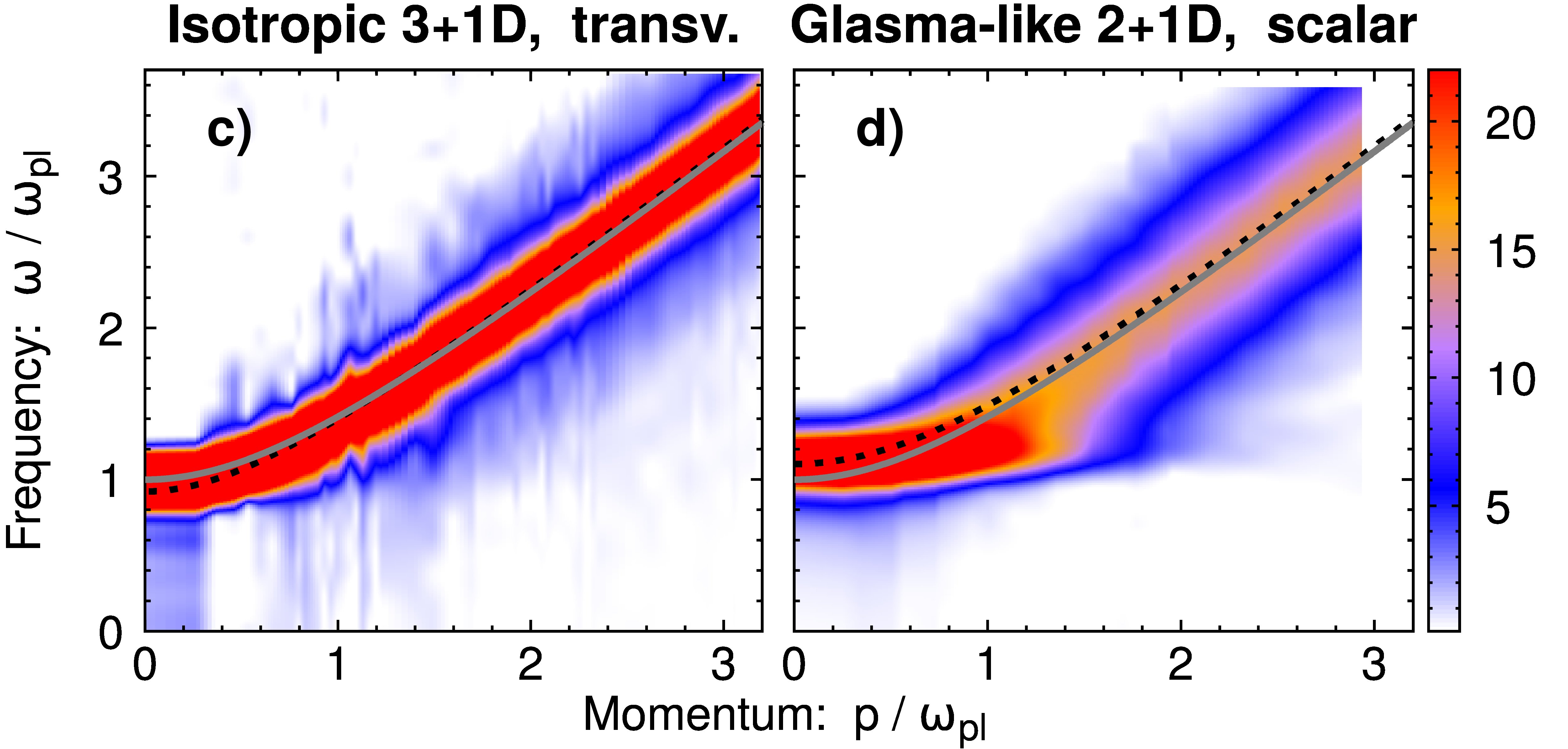}
	\caption{
	The extracted and normalized transverse statistical correlation function $\langle EE \rangle_{T}(t,\omega,p)/\langle EE \rangle_{T}(t,\Delta t {=} 0,p)$	is shown for a) the $2+1$ dimensional theory, b) for the Glasma-like theory and c) for an isotropic $3+1$ dimensional system. Additionally, we show the scalar correlator $\langle \pi\pi \rangle(t,\omega,p)/\langle \pi\pi \rangle(t,\Delta t {=} 0,p)$ of the Glasma-like system in d). 
	Note that all amplitudes have been cut off at the same fixed value $22/\Q$ corresponding to the red region.
	For comparison, we added the HTL dispersion relations $\omega_{T/\phi}^\HTL(p)$ as black dashed lines and a relativistic dispersion $\omega_{\rm{rel}}(p)=\sqrt{\wplas^2+p^2}$ as a continuous gray line with the extracted values for $\wplas$ as in the text.
	}
	\label{fig_2D_trans_ddF}
\end{figure}

\begin{figure}[t]
	\centering
	\includegraphics[scale=0.06]{\pToFigs/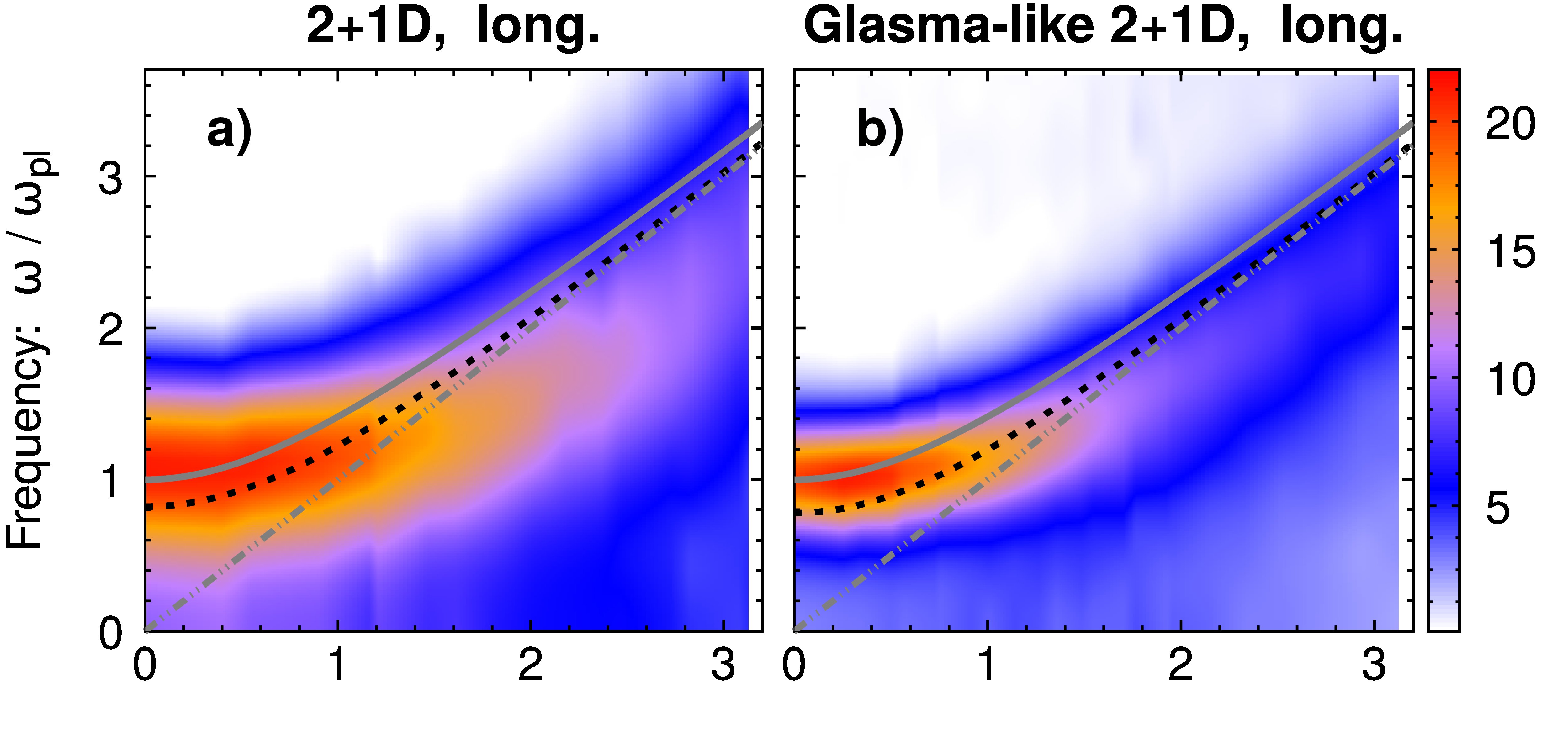}
    \includegraphics[scale=0.06]{\pToFigs/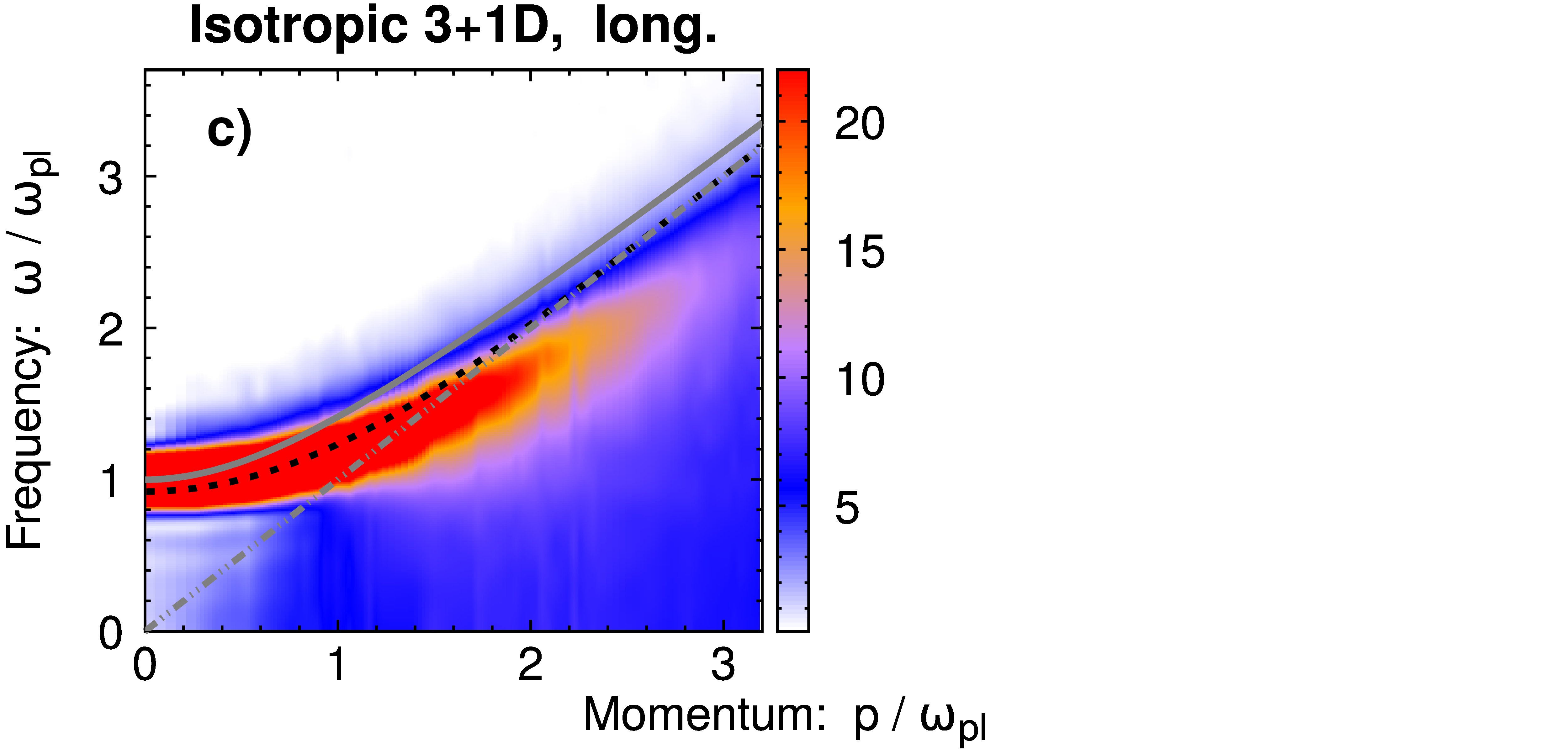}
	\caption{The normalized longitudinal statistical correlator $\langle EE \rangle_{L}(t,\omega,p)/\langle EE \rangle_{L}(t,\Delta t {=} 0,p)$	is plotted for a) the $2+1$ dimensional theory, b) for the Glasma-like theory and c) for an isotropic $3+1$ dimensional system.
	We also included the HTL dispersion relation $\omega_L^\HTL(p)$ as black dashed lines, a relativistic dispersion $\omega_{\rm{rel}}(p)$ as a continuous gray curve and the free dispersion $\omega=p$ as gray dash-dotted lines.}
	\label{fig_2D_long_ddF}
\end{figure}

\section{Numerical results}
\label{sec:results}

In this section, we present our numerical results for the excitation spectrum of non-equilibrium 2+1D theories by extracting spectral and statistical correlation functions. We briefly discuss some results in (classical) thermal equilibrium in Appendix~\ref{app_class}. 

For our simulations, we use the lattice size $1024^2$ and lattice spacing $\Q a_s = 1/8$ for the 2+1D system and $\Q a_s = 1/4$ on a $512^2$ lattice for the Glasma-like 2+1D system. We showed in \cite{Boguslavski:2019fsb} for equal-time correlation functions that for the considered initial conditions, these discretization parameters are sufficient to avoid lattice artifacts. Our resulting correlation functions will be compared with those from isotropic 3+1D systems computed in \cite{Boguslavski:2018beu} with $\Q a_s = 0.7$ on a $256^3$ lattice, which has been demonstrated not to show any significant lattice artifacts for this choice of parameters. 

While respecting the Gauss law constraint to almost machine precision, the linearized fluctuations on top of the inhomogenous, time-dependent, background field do not have conservation laws of quantities like energy, momentum and angular momentum. Thus, there is no guarantee that they would not grow strongly with time, leading to increasing error bands. Indeed, we observe such a growth, more strongly in two than in three spatial dimensions. This limits in practice our ability to extract the spectral function in the time domain. In practice we do not extend our extraction of the spectral function for 2+1D theories to more than $\Delta t \leq \dtmax = 80/Q$. No such restrictions apply for statistical correlation functions, which allows us to simulate to much later times. While it suffices to use $\dtmax = 80/\Q$ also for transverse and longitudinal statistical correlators in 2+1D theories, unless stated otherwise, we employ $\dtmax = 200/\Q$ for the scalar statistical correlator of the Glasma-like theory to correctly capture the long-lived scalar excitations at low momenta.

All the results in the figures are shown at $\Q t = 500$ for the 2+1D systems and $\Q t = 1500$ for the 3+1D theory, unless stated otherwise. From our earlier papers \cite{Boguslavski:2018beu,Boguslavski:2019fsb} we see that the systems at these times are well in the universal scaling regime discussed in \se\ref{sec:self-sim}. We will discuss how the correlators depend on time $t$ in \se\ref{sec:disp_gamma}.

Correlation functions as functions of frequency or relative time for fixed momenta are shown with error bars. These result from statistical averaging of the correlation functions over 320 - 400 configurations in frequency space or in the relative time domain, respectively.

\subsection{Transverse, longitudinal and scalar correlations}
\label{sec:ColorPlots}

Figure \ref{fig_2D_trans_ddF} shows the extracted normalized statistical correlation functions 
$\langle EE \rangle_{\alpha}(t,\omega,p) /$ $ \langle EE \rangle_{\alpha}(t,\Delta t {=} 0,p)$ 
as functions of frequency and momentum for transverse excitations in 2+1D theory (a), Glasma-like 2+1D (b), and isotropic 3+1D (c), and for scalar excitation in the Glasma-like 2+1D theory (d).

All frequencies and momenta of the correlators are normalized using the mass $\wplas$, which is the plasmon frequency of gluonic polarizations. While there are multiple ways of determining $\wplas$ \cite{Lappi:2016ato,Lappi:2017ckt}, we extract it from the dotted spectral function $\dot{\rho}_T(t,\omega,p {=} 0)$ at vanishing momentum using a (Gaussian) fit function, as explained in \se\ref{sec:peaks}. 
At the considered times, the values of the plasmon frequency read 
$\wplas^{\tDTwMi} = 0.12\,\Q$ for the 2+1D theory, $\wplas^{\tDThrMi} = 0.2\,\Q$ for the Glasma-like 2+1D theory and $\wplas^{\tThrD} = 0.13\,\Q$ for the 3+1D theory. 
The continuous gray lines in \fig\ref{fig_2D_trans_ddF} correspond to a relativistic dispersion $\omega_{\rm{rel}} = \sqrt{\wplas^2 + p^2}$ with the numerically extracted $\wplas$. We show the transverse and scalar HTL dispersion relations $\omega_{T/\phi}^\HTL(p)$ from \eq\eqref{eq_HTL_trans} and \eq\eqref{eq_HTL_scalar} as black dashed lines. Note that the relativistic dispersion is an ad hoc ansatz with the parameter $\wplas$ extracted by fitting the data. 
The HTL dispersion relation, in contrast, has a functional form determined by the leading order HTL calculation, and a mass scale determined by the distribution function $f(t,p)$ using \eq\eqref{eq_mHTL_selfconst}. For the HTL mass, the resulting values are $m_\HTL^{\tDTwMi} = 0.14\,\Q$, $m_\HTL^{\tDThrMi} = 0.21\,\Q$ and $m_\HTL^{\tThrD} = 0.15\,\Q$ for the considered parameters and times of the different theories.

When considering the correlation functions in \fig\ref{fig_2D_trans_ddF} for fixed momentum, one observes an excitation peak as a function of frequency for each of the different non-Abelian gauge theories and their transverse gluonic or scalar contributions. The additionally shown relativistic dispersion $\omega_{\rm{rel}}(p)$ and HTL dispersion relations lie well within the width of the peak. 
Interestingly, the scalar excitation shows some deviations from a relativistic dispersion, which actually is the leading order HTL prediction \eqref{eq_HTL_scalar}. These deviations are at their strongest at $p \sim \wplas$. 
However, at larger momenta $p \gg \wplas$ the dispersion relation coincides with its HTL prediction, which also agrees with the transverse HTL dispersion for large momenta. As we will see in \se\ref{sec_scalar}, also the shape of the spectral function agrees between transverse and scalar polarizations at higher momenta. 

An important difference to the expected form of the excitations within the hard-loop framework lies in the width of the peaks (considered as functions of $\omega$), which in hard loop theory is a next-to-leading order effect and is thus supposed to be small. We also observe that for gluonic excitations the peaks in 2+1D theories are much wider than in 3+1D or for low-momentum scalar excitations, whose width seems to be comparable to the gluonic width in 3+1D.%
\footnote{Note that in the lower panels of \fig\ref{fig_2D_trans_ddF}, the amplitudes of the 3+1D and scalar excitations have been cut off at the same value as for the transverse 2+1D excitations in the upper panels to enable a direct comparison. However, their amplitudes reach values around $80$, such that the peak width visible in the figure appears larger than in reality.}
Transverse gluonic excitations of 2+1D theories at low momenta $p \lesssim \wplas$ also show sizeable contributions all the way to $\omega=0$. We will discuss this observation in more detail below in Sec.~\ref{sec:smallomega}.

Figure~\ref{fig_2D_long_ddF} shows the normalized longitudinal statistical correlator \\
$\langle EE \rangle_{L}(t,\omega,p) / \langle EE \rangle_{L}(t,\Delta t {=} 0,p)$ 
of 2+1D (a), Glasma-like 2+1D (b), and isotropic 3+1D theory (c) for comparison. The gray dash-dotted lines correspond to the free dispersion $\omega=p$, the gray continuous lines to $\omega_{\rm{rel}}(p)$ and the black dashed curves to the HTL dispersion relations $\omega_L^\HTL(p)$ from \eq\eqref{eq_HTL_long}. 
One observes that the peaks at low momenta $p \lesssim \wplas$ have a similar width and magnitude as the corresponding transversely polarized gluonic peaks of \fig\ref{fig_2D_trans_ddF}, and agree with the longitudinal HTL dispersion reasonably well. At all momenta, one finds that the longitudinal correlations involve a finite valued continuum for $\omega \lesssim p$, which corresponds to the Landau cut contribution. Different from transverse dotted spectral functions, the quasiparticle peaks of longitudinal correlators are strongly suppressed at higher momenta $p \gtrsim \wplas$ and Landau damping becomes the dominant contribution, as visible in the figure. 

These properties will be studied in more detail by comparing with HTL calculated curves in the following subsection. Note that while we have shown the statistical correlation functions $\langle EE \rangle_{\alpha}(t,\omega,p)$ here, they appear to be related to the respective dotted spectral functions $\dot{\rho}_{\alpha}(t,\omega,p)$ by the generalized fluctuation-dissipation relation \eqref{eq_ddF_drho_rel} even in our far-from-equilibrium situation, as will also be shown in the following. Therefore, the discussions of this subsection are also valid for the dotted spectral functions in the considered theories.

\begin{figure}[t]
	\centering
	\includegraphics[scale=0.8]{\pToFigs/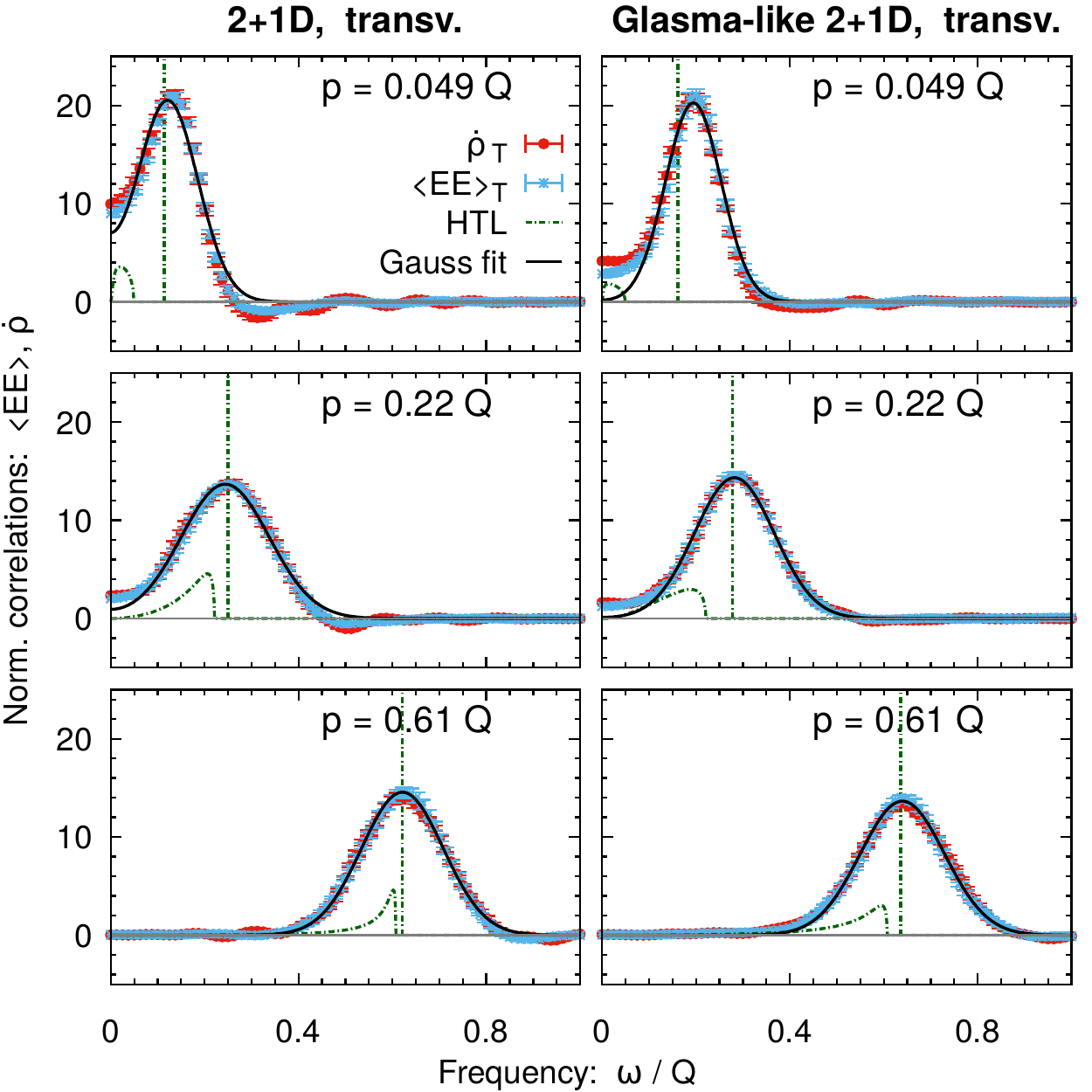}
	\caption{The transverse normalized statistical correlator $\langle EE \rangle_{T}(t,\omega,p)/\langle EE \rangle_{T}(t,\Delta t {=} 0,p)$ and dotted spectral function $\dot{\rho}_{T}(t,\omega,p)$ for both $2+1$-dimensional theories at different momenta as functions of frequency. 
	They lie on top of each other within numerical precision, demonstrating the validity of the generalized fluctuation-dissipation relation \eqref{eq_ddF_drho_rel}.
	The HTL expressions for $\omega \rho$ are shown as green dashed lines while black lines denote fits using the Gaussian distribution in \eqref{eq:gauss_dist}.
	}
	\label{fig_2D_trans_ddF_differentPeaks}
\end{figure}

\begin{figure}[t]
	\centering
	\includegraphics[scale=0.8]{\pToFigs/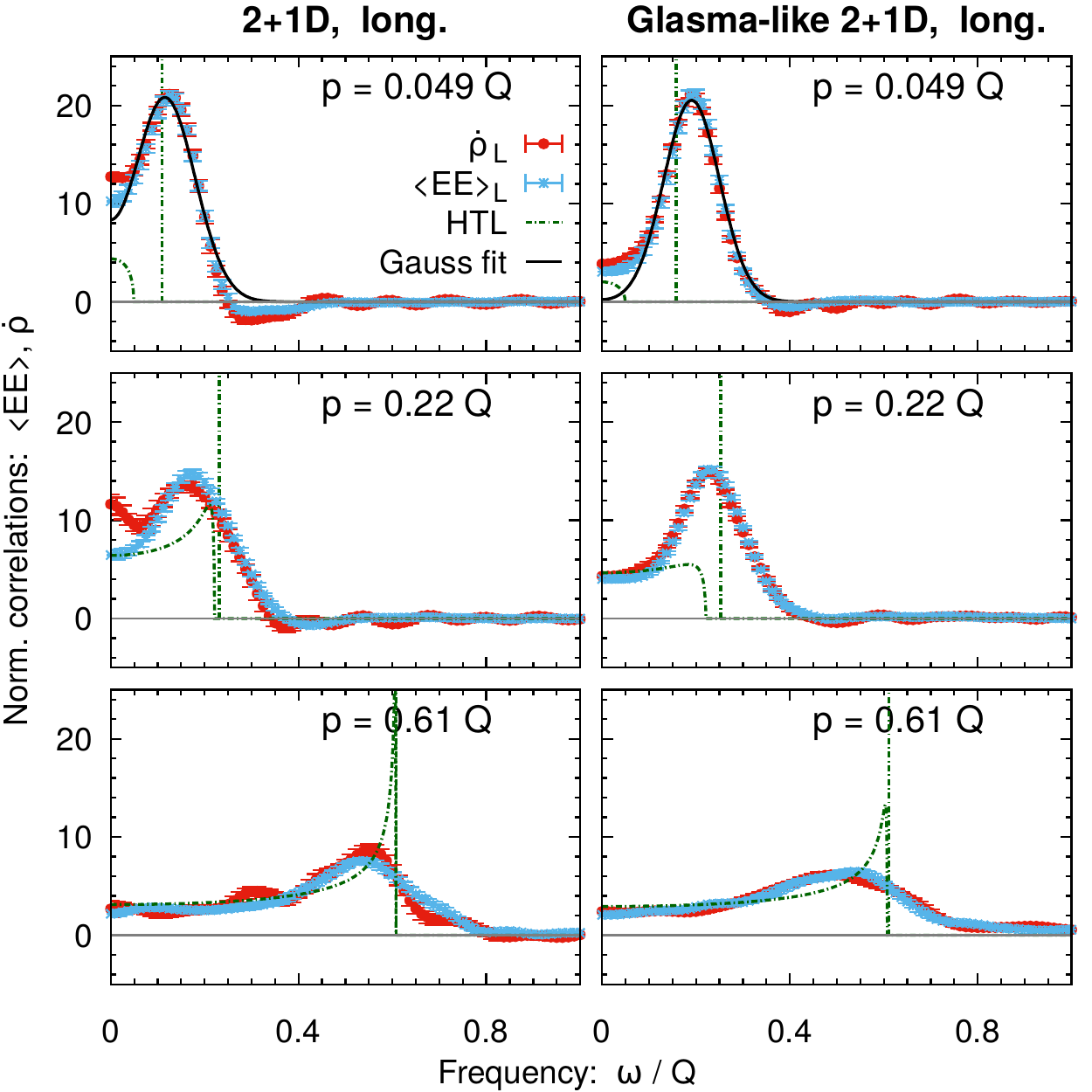}
	\caption{
	The longitudinal normalized statistical correlator $\langle EE \rangle_{L}(t,\omega,p)/\langle EE \rangle_{L}(t,\Delta t {=} 0,p)$ and normalized dotted spectral function $\dot{\rho}_{L}(t,\omega,p)/\dot{\rho}_{L}(t,\Delta t {=} 0,p)$ for both 2+1D theories for the same momenta and with analogous HTL and fitting curves as in \fig\ref{fig_2D_trans_ddF_differentPeaks}.
	For $p = 0.61\,\Q$ we used the smaller time window $\dtmax = 60/\Q$ to reduce ringing of the dotted spectral function.
	}
	\label{fig_2D_trans_ddF_differentPeaks_long}
\end{figure}

\subsection{Comparison of spectral and statistical correlation functions}
\label{sec_gen_fluct_dis}

In the isotropic 3+1D theory, we observed in \cite{Boguslavski:2018beu} that the spectral and statistical correlation functions in our considered far-from-equilibrium system are related by the generalized fluctuation-dissipation relation 
\begin{align}
 \label{eq_ddF_drho_rel}
 \frac{\langle EE \rangle_{\alpha}(t,\omega,p)}{\langle EE \rangle_{\alpha}(t,\Delta t {=} 0,p)} \approx \frac{\dot{\rho}_{\alpha}(t,\omega,p)}{\dot{\rho}_{\alpha}(t,\Delta t {=} 0,p)} \,.
\end{align}
We will show here that this relation also holds in 2+1D theories for each polarization $\alpha = T,L,\phi$. The respective equal-time dotted spectral functions are determined by the sum rules \eqref{eq_sum_rules_2D_T}-\eqref{eq_sum_rules_2D_phi} while the equal-time statistical correlators $\langle EE \rangle_{\alpha}(t,\Delta t {=} 0,p)$ are extracted numerically. 
The generalized fluctuation-dissipation relation \eqref{eq_ddF_drho_rel} implies that the normalized statistical and dotted spectral functions agree quantitatively, which is nontrivial out of equilibrium. In thermal equilibrium both correlations are connected by the fluctuation-dissipation relation, which for low momenta $p \ll T$ below the temperature reads $\langle EE \rangle_{\alpha}(\omega,p) = T\,\dot{\rho}_{\alpha}(\omega,p)$. In contrast, out of equilibrium their connection may be much more complicated or even absent, as observed for scalar theories \cite{PineiroOrioli:2018hst,Boguslavski:2019ecc,Shen:2019jhl}. 
The HTL framework predicts a generalized fluctuation-dissipation relation \cite{Arnold:2002zm} with a time-dependent effective temperature $T_{\text{eff}}(t) = \langle EE \rangle_{\alpha}(t,\Delta t {=} 0,p)/\dot{\rho}_{\alpha}(t,\Delta t {=} 0,p)$
\begin{align}
\label{eq:ddF_drho_rel_HTL}
 \langle EE \rangle_{\alpha}(t,\omega,p) \approx T_{\text{eff}}(t)\,\dot{\rho}_{\alpha}(t,\omega,p)
\end{align}
to hold for sufficiently small momenta $p \ll \Lambda(t)$. Here $\Lambda(t)$ is a time-dependent hard scale that dominates the energy density and plays the role of a temperature scale for momenta in the non-equilibrium setting. However, we emphasize that we observe the relation \eqref{eq_ddF_drho_rel} to be satisfied in general. This also includes higher momenta $p \gtrsim \Lambda(t)$, where $\langle EE \rangle_{\alpha}(t,\Delta t {=} 0,p)/\dot{\rho}_{\alpha}(t,\Delta t {=} 0,p)$ are non-constant functions of $t$ and $p$.

The validity of the generalized fluctuation-dissipation relation \nr{eq_ddF_drho_rel} in 2+1D non-Abelian plasmas is one of the main results of our work. It is demonstrated in \figs\ref{fig_2D_trans_ddF_differentPeaks} and \ref{fig_2D_trans_ddF_differentPeaks_long}, which show the normalized statistical and spectral correlation functions for three different momenta for the 2+1D (left panels) and for the Glasma-like theory (right panels) for transverse and longitudinal polarizations, respectively. 
The normalized statistical correlation functions (blue points) precisely overlap with the corresponding dotted spectral functions (red points), up to residual ringing of the dotted spectral functions that results from the finite window size and statistical fluctuations of the Fourier transform. Thus, the figures confirm the generalized fluctuation-dissipation relation given by \eq\eqref{eq_ddF_drho_rel}.

We also observe in these figures that the width and shapes of the correlation functions look very similar in both the 2+1D and Glasma-like 2+1D theories for transverse and longitudinal excitations, when plotted as functions of $\omega/Q$. Note that the scale $Q$ is related to the energy density \emph{per polarization state}. Thus, adding the scalar degrees of freedom to the system without changing the distribution $f(p)$ of the gauge particles does not seem to modify the shape or the width. In contrast to the width of the peak, the mass in the Glasma-like system is always larger, as seen particularly well by the shift of the peak at lower momenta. 
Thus, the squared plasmon mass is proportional to the number of polarization states, as expected from HTL in \eq\eqref{eq_mHTL_selfconst}. Taking together these two observations seems to indicate that, to a first approximation, the scalar sector contributes to the mass but does not contribute to the damping of gluonic excitations. 

If we plotted  the correlation functions at the same $p/\wplas$ as functions of $\omega/\wplas$, i.e., normalized by the plasmon mass of the corresponding theory instead of an energy-related momentum scale, the peak positions would overlap by construction. However, the peaks in the 2+1D Glasma-like system would appear more narrow than in the 2+1D theory. This effect, which we have already seen in \figs\ref{fig_2D_trans_ddF} and \ref{fig_2D_long_ddF}, means that in the Glasma-like theory the quasiparticle excitations are slightly longer lived than in pure 2+1D.

We now compare the correlators to the transverse and longitudinal spectral functions calculated in HTL \eqref{eq:rhoHTL_decomposition}, which consist of a Landau cut contribution and a quasiparticle delta peak. They are shown as green dashed lines in \figs\ref{fig_2D_trans_ddF_differentPeaks} and \ref{fig_2D_trans_ddF_differentPeaks_long}. 
Importantly, different from 3+1D gluonic plasmas, the extracted dotted spectral functions are so broad that an accurate distinction between Landau cut region and an excitation peak becomes difficult for both polarizations. Therefore, the HTL expressions mostly do not provide a good description of the nonperturbative simulation results of the 2+1D systems. 

One exception is the large-momentum region $p \gtrsim m_\HTL$ of the longitudinally polarized dotted spectral function, which concerns the central and lower panels of \fig\ref{fig_2D_trans_ddF_differentPeaks_long}. According to HTL expectations that are discussed in \app\ref{app_HTL2D}, the Landau damping term is expected to dominate the spectral function for these momenta. We indeed observe that the HTL Landau damping curves for low frequencies $\omega \ll p$ agree well with the numerical data. This observation, which validates the HTL formulas for this specific case, is nontrivial, since the HTL framework is not expected to work well for the 2+1D theories due to the importance of the missing soft-soft interactions, as discussed in \se\ref{sec:HTL}. 

Note that the sharp peaks of the HTL Landau contributions around $\omega \sim p$ are smoother in the numerically extracted correlators. We have reported of a similar smoothening of the Landau cut region for 3+1D plasmas in \cite{Boguslavski:2018beu}, which may result from including more mode interactions than considered in HTL at leading order.

\begin{figure}[t]
	\centering
	\includegraphics[scale=0.8]{\pToFigs/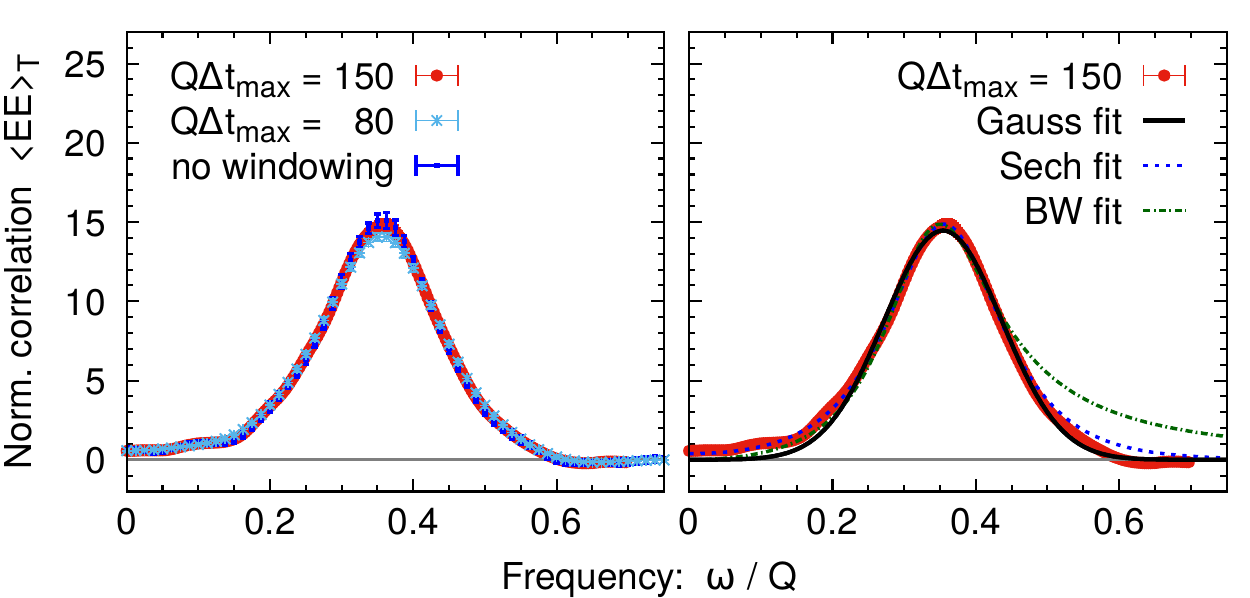}
	\caption{The normalized transverse statistical correlator $\langle EE \rangle_{T}(t,\omega,p)/\langle EE \rangle_{T}(t,\Delta t {=} 0,p)$ of Glasma-like 2+1D system for momentum $p = 0.3\,\Q$. 
	In the left panel, different ways to compute the Fourier transform are compared, which are seen to agree well. 
	In the right panel, we compare different fit functions for the peak shape, as explained in the main text.
	}
	\label{fig_2D_trans_ddF_params}
\end{figure}

\subsection{Shape of the peaks}
\label{sec:peaks}

Let us now discuss the shape of the excitations in more detail. Computations within the HTL formalism (\se\ref{sec:HTL}) suggest that the normalized correlation functions can be written as a sum 
\begin{align}
 \frac{\dot{\rho}_{\alpha}(t,\omega,p)}{\dot{\rho}_{\alpha}(t,\Delta t {=} 0,p)} \approx h(\omega, p) + \frac{\omega\,\rho_{\alpha}^{\mrm{Landau}}(t,\omega,p)}{\dot{\rho}_{\alpha}(t,\Delta t {=} 0,p)}
\end{align}
with an excitation peak $h(\omega, p)$ that originates from a pole at frequency $\pm \omega_\alpha(p)$ with residue $Z_{\alpha}(p)$ and becomes a Delta function at leading order in HTL. A finite peak width should thus be a subleading-order effect. Since the peak results from poles in the retarded propagator (\eqs\eqref{eq_GR_HTL_T}-\eqref{eq_GR_HTL_N}), its shape could be expected to be of Breit-Wigner form, which becomes a Lorentzian distribution for narrow peaks. A Lorentzian shape for the excitation peaks has been found in 3+1D gluonic systems \cite{Boguslavski:2018beu} and would usually correspond to quasiparticle excitations.

The peak form is investigated in \fig\ref{fig_2D_trans_ddF_params}. In the right panel, the normalized transverse statistical correlator for the larger time window $\Q \dtmax = 150$ is shown in red for the Glasma-like theory at momentum $p = 0.3\,\Q$. It is compared to fits $h(\omega, p)$ that are modeled as a Gaussian, hyperbolic secant and Breit-Wigner distribution, respectively, given by
\begin{align}
\label{eq:gauss_dist}
 h_{\text{Gauss}} &= \frac{A}{\gamma} \sqrt{\frac{\pi}{2}}\,  \text{exp}\!\left[-\frac{1}{2}\left(\frac{\omega-\omega_{R}}{\gamma}\right)^2\right] + [\omega_{R}\mapsto-\omega_{R}] \\
 h_{\text{Sech}} &= \frac{A}{\gamma}\frac{\pi}{2}\, \text{sech}\!\left[\frac{\pi}{2}\,\frac{\omega-\omega_R}{\gamma}\right] + [\omega_{R}\mapsto-\omega_{R}] \\
 h_{\text{BW}} &= \frac{a\,\omega^2}{(\omega^2-\omega_R^2-\gamma^2)^2+4\omega^2\gamma^2}
\end{align}
with $\text{sech}(x) = 1/\text{cosh}(x)$. Here $[\omega_{R}\mapsto-\omega_{R}]$ denotes the previous term with the indicated substitution.%
\footnote{All parameters, which are the dispersion relations $\omega_D$ and the width $\gamma$ are momentum and time dependent and will be discussed in \se\ref{sec:disp_gamma}. The amplitude $A$ corresponds to the residue of the quasiparticles and is close to unity for transverse excitations.}
We neglect the Landau cut contribution in the fitting procedure because it is subleading and does not describe our data well at frequencies $\omega \sim p$, which lie within the width of the excitation peak.
Comparing the different fits in the plot, one observes that the Breit-Wigner distribution fails to describe the peak faithfully, which is particularly well visible at the right tail. 
In contrast, the shape of the peak is well described by the non-Lorentzian distributions $h_{\text{Gauss}}$ and $h_{\text{Sech}}$. Indeed, we have included corresponding fits to our data of a Gaussian distribution $h_{\text{Gauss}}$ 
for different momenta and polarizations already in the \figs\ref{fig_2D_trans_ddF_differentPeaks} and \ref{fig_2D_trans_ddF_differentPeaks_long} as black lines. The fits are seen to agree well with the shape of the dotted spectral functions for all momenta, where such a comparison is sensible. Note that for longitudinal spectral functions, quasiparticle excitations are expected to be suppressed at higher momenta $p \gtrsim m_D$. Therefore, we only included fits for small momenta $p \ll m_D$ in \fig\ref{fig_2D_trans_ddF_differentPeaks_long}.

We have checked that the non-Lorentzian shape is not an artefact of a finite time window in the Fourier transform or of signal processing techniques. In the left panel of \fig\ref{fig_2D_trans_ddF_params} we show the same correlation function as in the right panel for the time windows $\Q \dtmax = 150$ and $\Q \dtmax = 80$ with Hann windowing as well as for $\Q \dtmax = 80$ without windowing, which implies $h(\Delta t) = 1$ in \eq\eqref{eq:FTrafo}. One observes that all curves accurately agree within uncertainties.

This non-Lorentzian peak, thus, appears to be a distinct feature of 2+1D theories in contrast to 3+1D plasmas, where Lorentzian excitations emerge instead \cite{Boguslavski:2018beu}. A similar non-Lorentzian shape has been encountered in single-component non-relativistic and relativistic scalar models at low momenta \cite{PineiroOrioli:2018hst,Boguslavski:2019ecc}. It was also used to distinguish the corresponding excitations from the usual Lorentzian peaks that dominate in $O(N)$-symmetric scalar models for a large number of components $N \gg 1$ \cite{Boguslavski:2019ecc}. For the  2+1D gauge theories  considered here, the origin for the observed non-Lorentzian form is currently unknown.

\subsection{Scalar excitation}
\label{sec_scalar}

\begin{figure}[t]
	\centering
	\includegraphics[scale=0.8]{\pToFigs/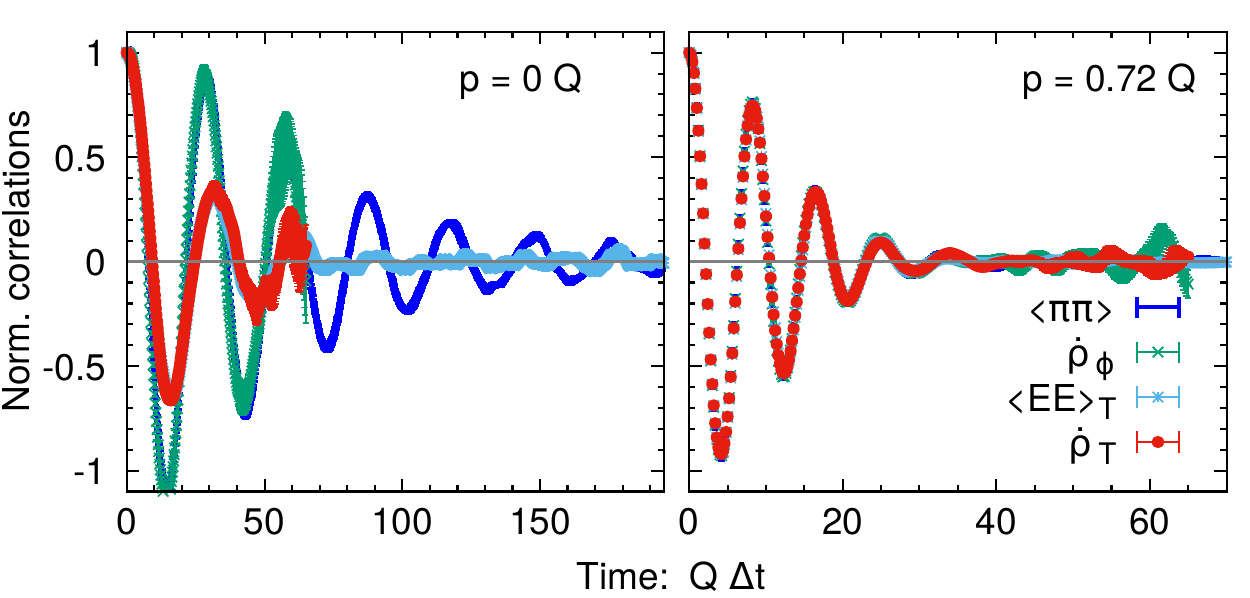}
	\includegraphics[scale=0.8]{\pToFigs/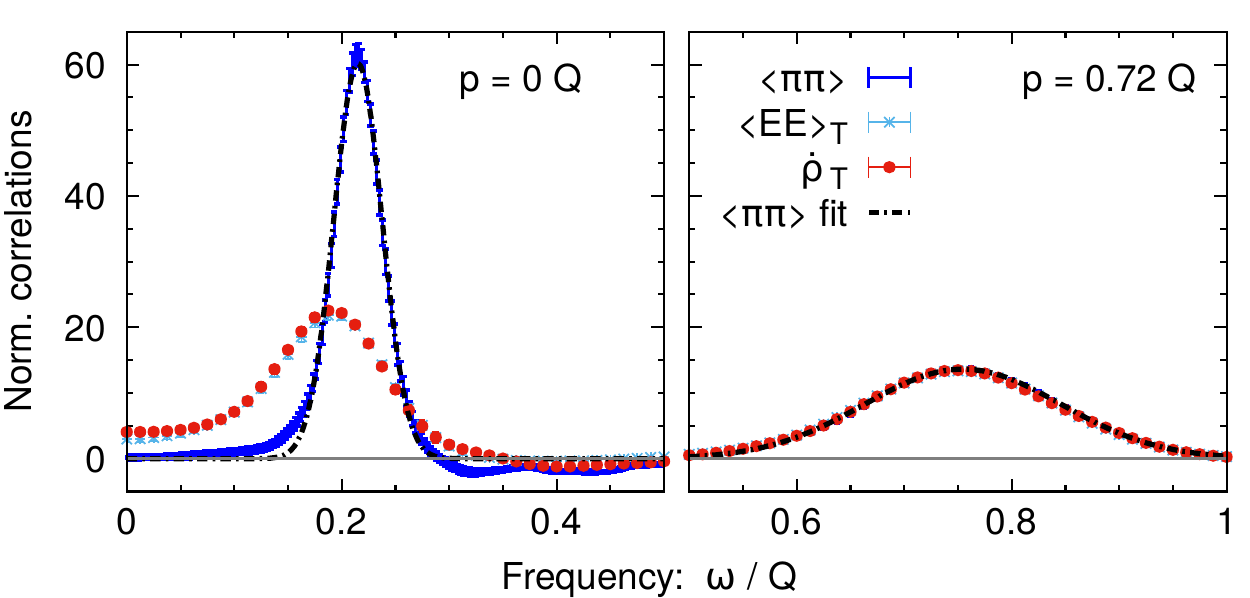}
	\caption{
	Different scalar and transverse correlators of the Glasma-like 2+1D theory for momenta $p = 0\,\Q$ (left) and $p = 0.72\,\Q$ (right). The scalar correlations are seen to also satisfy the generalized fluctuations-dissipation relation and even to agree with transverse correlations at high momenta, while differing at low momenta.
	{\bf Top:} The correlation functions $\langle \pi\pi \rangle / \langle \pi\pi \rangle(\Delta t {=} 0,p)$, $\dot{\rho}_\phi$, $\langle EE \rangle_T / \langle EE \rangle_T(\Delta t {=} 0,p)$ and $\dot{\rho}_T$ as functions of time $\Delta t$. Note that in the left panel the range of $\Delta t$ is longer.
	{\bf Bottom:} The same correlators as in the top row but now in the frequency domain $\omega$. The scalar dotted spectral function $\dot{\rho}_\phi$ is not shown because $\Q \dtmax$ of at least $200$ would be necessary while the correlator becomes very noisy for $\gtrsim 80$. Gaussian fits to the scalar correlator $\langle \pi\pi \rangle$ are shown as black dashed lines.
	}
	\label{fig_2D_ET_dt}
\end{figure}

We can also compare correlation functions in the (relative) time domain. Consistently with \eq\eqref{eq_ddF_drho_rel}, the generalized fluctuation-dissipation relation 
\begin{align}
 \label{eq_ddF_drho_rel_dt}
 \frac{\langle EE \rangle_{\alpha}(t,\Delta t,p)}{\langle EE \rangle_{\alpha}(t,\Delta t {=} 0,p)} \approx \frac{\dot{\rho}_{\alpha}(t,\Delta t,p)}{\dot{\rho}_{\alpha}(t,\Delta t {=} 0,p)} \,,
\end{align}
is also satisfied in $\Delta t$, where $\alpha$ denotes the polarizations $T,L,\phi$. This is demonstrated in the upper panels of \fig\ref{fig_2D_ET_dt} for the 2+1D Glasma-like theory, where the normalized statistical $\langle EE \rangle_{\alpha}(t,\Delta t,p)/\langle EE \rangle_{\alpha}(t,\Delta t {=} 0,p)$ and (dotted) spectral correlation functions $\dot{\rho}_{\alpha}$ are seen to nicely coincide for scalar and transverse polarizations separately. Interestingly, both polarizations even agree with each other at high momenta $p \gg m_D$ (right panels), while they are seen to differ at lower momenta $p \lesssim m_D$ (left panels). 

\begin{figure}[t]
	\centering
	\includegraphics[scale=0.8]{\pToFigs/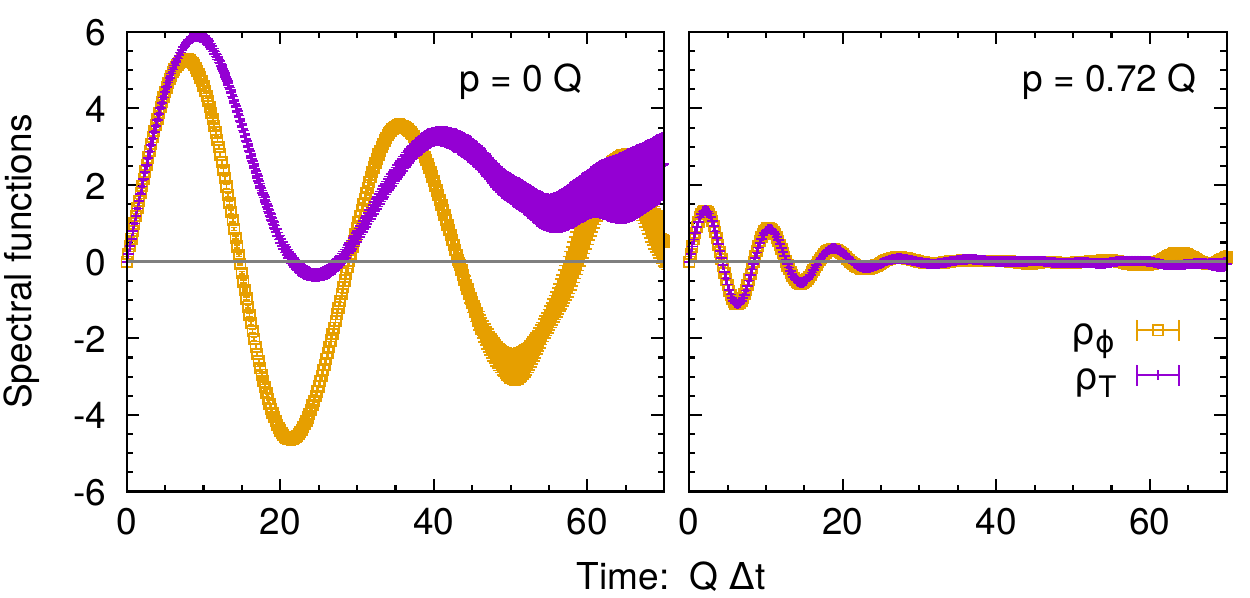}
	\caption{
	The actual scalar and transverse spectral functions $\rho_\phi$ and $\rho_T$ of the Glasma-like 2+1D theory for momenta $p = 0\,\Q$ (left) and $p = 0.72\,\Q$ (right) as functions of relative time $\Delta t$. Different from the dotted spectral function $\dot{\rho}_T$ in \fig\ref{fig_2D_ET_dt}, $\rho_T$ is seen to oscillate around a non-zero value in the left panel.
	}
	\label{fig_2D_rhoT_dt}
\end{figure}

\begin{figure}[t]
	\centering
	\includegraphics[scale=0.8]{\pToFigs/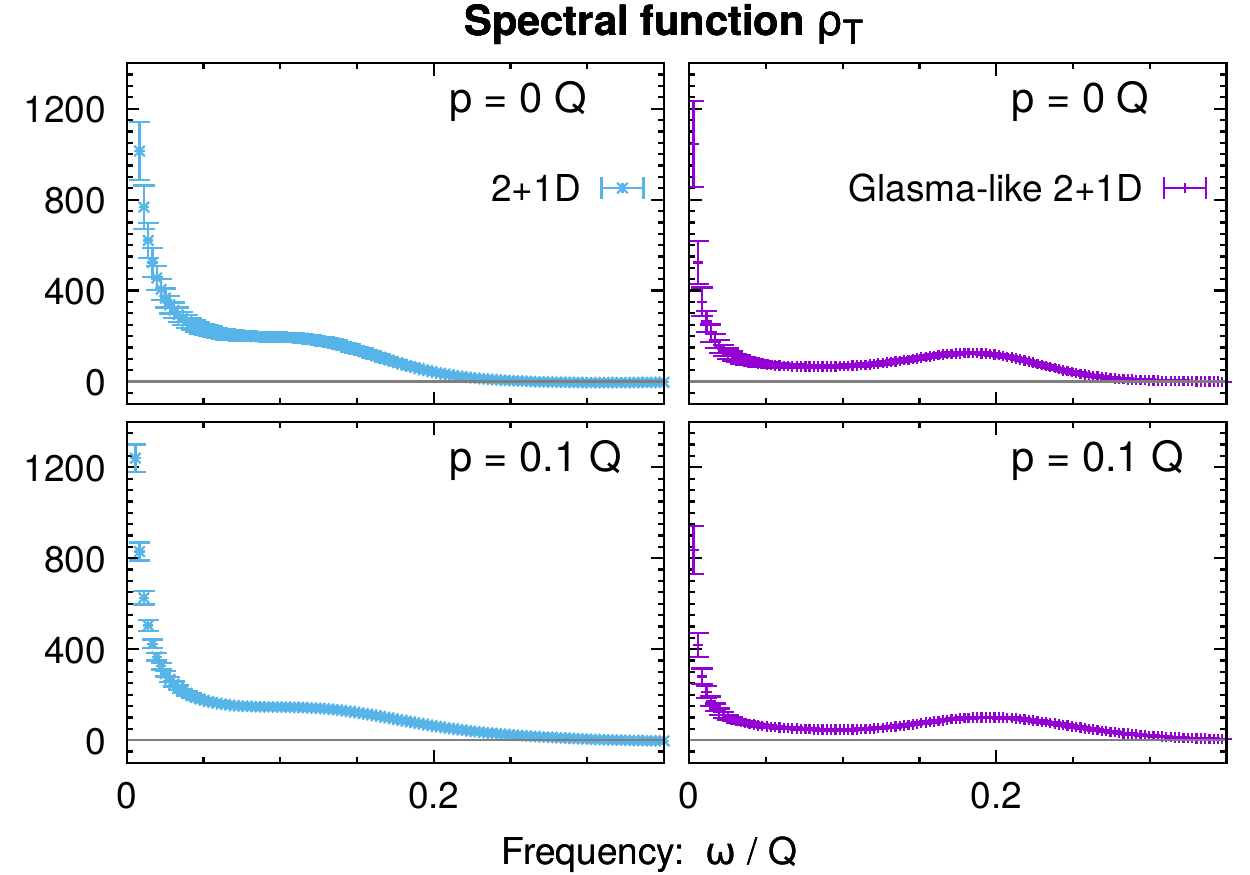}
	\caption{
	Transverse spectral function $\rho_T$ computed as $\dot{\rho}_T/\omega$ for small momenta $p=0\,\Q$ and $p=0.1\,\Q$ as a function of $\omega$ for 2+1D (left) and Glasma-like 2+1D (right) theories. Due to its symmetry, the spectral function satisfies $\rho_T(t,\omega{=}0)=0$ identically while it follows $\rho_T \sim 1/\omega$ for $\omega \to 0$ due to the finite values of $\dot{\rho}_T$.
	}
	\label{fig_2D_rhoT_w}
\end{figure}

This observation implies that the distinction between transverse and scalar excitations is only relevant for low momenta.
We confirm this interpretation in the frequency domain in the lower panels of \fig\ref{fig_2D_ET_dt}. In the right panel we show the correlation functions for the same higher momentum and also include a (Gaussian) fit to the scalar correlator given by \eqref{eq:gauss_dist}. All correlators lie on top of each other, revealing the same shape, dispersion and width among transverse and scalar polarizations. In contrast, at the lowest momentum $p=0$, a clear difference is visible between the gluonic $\langle EE \rangle_T$ and the scalar correlation $\langle \pi\pi \rangle$. The scalar excitation is much more narrow, leading to longer-lived quasiparticles than for transverse gluonic excitations. We have observed this property also in \fig\ref{fig_2D_trans_ddF}d) for low momenta $p \lesssim m_D$, which is similar to the gluonic excitations in a spatially three-dimensional system. Furthermore, the peak position is shifted towards a slightly higher frequency. Interestingly, this agrees qualitatively with the HTL dispersion relations (\app\ref{app:spectral}), which predict that the peak positions of the gluonic and scalar excitations at $p=0$ are $\wplas$ and $m_D$, respectively, with $\wplas < m_D$.

\subsection{Low frequency behavior}
\label{sec:smallomega}

We now turn to the actual spectral functions $\rho_\alpha$, only having discussed their time derivative $\dot\rho_{\alpha}$ so far. For the scalar polarization, we see in \fig\ref{fig_2D_ET_dt} that $\dot\rho_{\phi}$ as a function of $\Delta t$ always oscillates around zero and has a narrow excitation peak in the frequency domain with a vanishing value for $\omega \to 0$. As visible in \fig\ref{fig_2D_rhoT_dt}, the scalar spectral function $\rho_\phi$ also oscillates around zero. 
This confirms that its Fourier transform $\rho_\phi(t,\omega,p)$ smoothly approaches zero as $\omega \to 0$. 

In contrast, the dotted spectral function of gluonic excitations $\dot\rho_{T}(t,\omega,p)$ is seen to behave differently than the scalar one. This is visible in frequency domain in \figs\ref{fig_2D_trans_ddF_differentPeaks} and \ref{fig_2D_ET_dt}, where the spectral function  approaches non-vanishing values for $\omega \to 0$ at low momenta $p \lesssim m_D$. As a consequence, the resulting spectral function shown in \fig\ref{fig_2D_rhoT_w} (computed as $\rho_{T} = \dot\rho_{T}/\omega$ as discussed earlier) behaves as $\sim 1/\omega$ for $\omega \to 0$. Thus, although $\rho_T(t,\omega{=}0)=0$ due to the odd symmetry, the spectral function actually behaves as $\rho_T(t,\omega) \sim 1/\omega$ for $\omega \to 0$, for both positive and negative $\omega$.
This is consistent with the behavior of $\rho_T$ as a function of $\Delta t$ for low momenta. This can be seen in the left panel of \fig\ref{fig_2D_rhoT_dt} where we observe that $\rho_T$ does not oscillate around zero but possibly approaches a nonzero value for $\Delta t \to \infty$. This behavior is surprising since, based on the analytical structure of the HTL self-energies, we would expect $\dot\rho_{T}$ and $\rho_{T}$ to smoothly vanish as $\omega \to 0$ and, consistently, $\rho_{T}(t,\Delta t,p)$ to oscillate around zero.

Such a behavior could, for instance, follow from having an additional excitation at $\omega \approx 0$ or a conserved quantity; however, we do not expect either here. 
Another possible explanation would be if it was related to time reversal non-invariance, since our system is evolving in time. To check this we have performed a similar calculation in the time reversal invariant case of classical thermal equilibrium in Appendix~\ref{app_class}. However, we also see a finite value for $\dot\rho_{T}(t,\omega{=}0,p)$ there for low momenta. Thus, the small violation of time reversal invariance in our system does not seem to be the explanation either. 

Another possibility is that this observed feature is associated with our gauge fixing procedure that is discussed in the end of \se\ref{sec:theory_corrs}. Assuming $\Delta t \ll t$, we fix to a Coulomb-type gauge at the time $t$, but the system does not exactly remain in Coulomb gauge at $t+\Delta t$. To check whether this approximation may lead to such an effect, one could perform the calculation fully in Coulomb gauge, which would avoid the mentioned approximation but require reintroducing a temporal component for the gauge potential. Moreover, it would be interesting to look at the heavy quark diffusion coefficient (as in \re\cite{Boguslavski:2020tqz}), which is a gauge invariant observable sensitive to the same physical scales. We leave these studies to a further paper. To summarize, we do not have a clear interpretation of the finite value of $\dot\rho$ at $\omega=0$.

\begin{figure}[t]
	\centering
	\includegraphics[scale=0.8]{\pToFigs/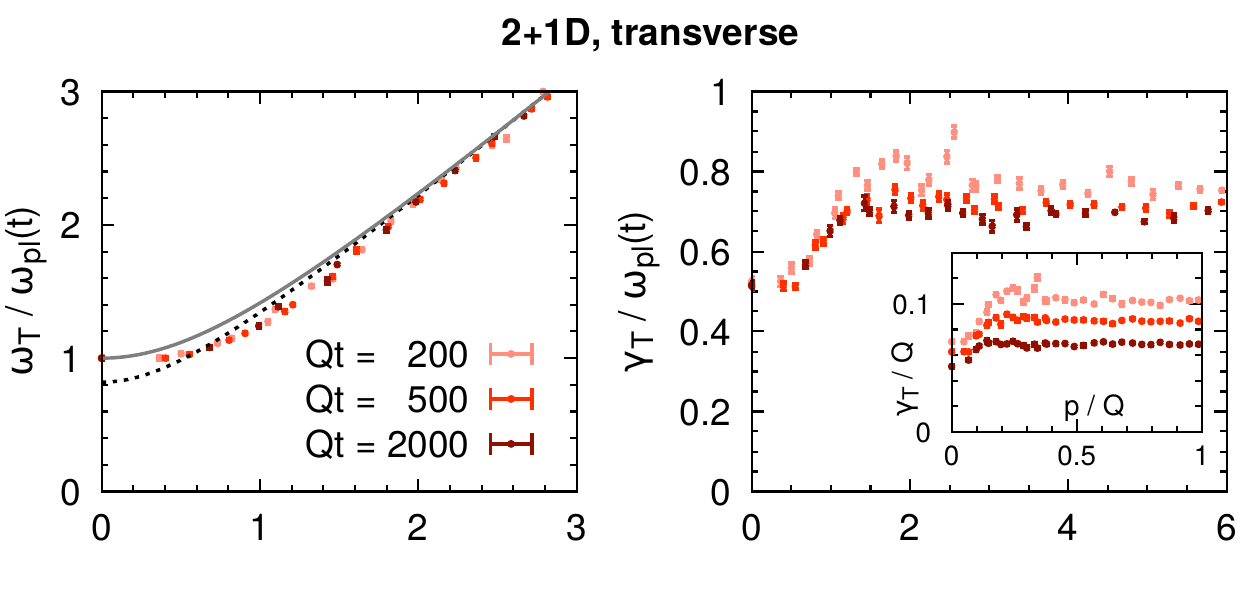}
	\includegraphics[scale=0.8]{\pToFigs/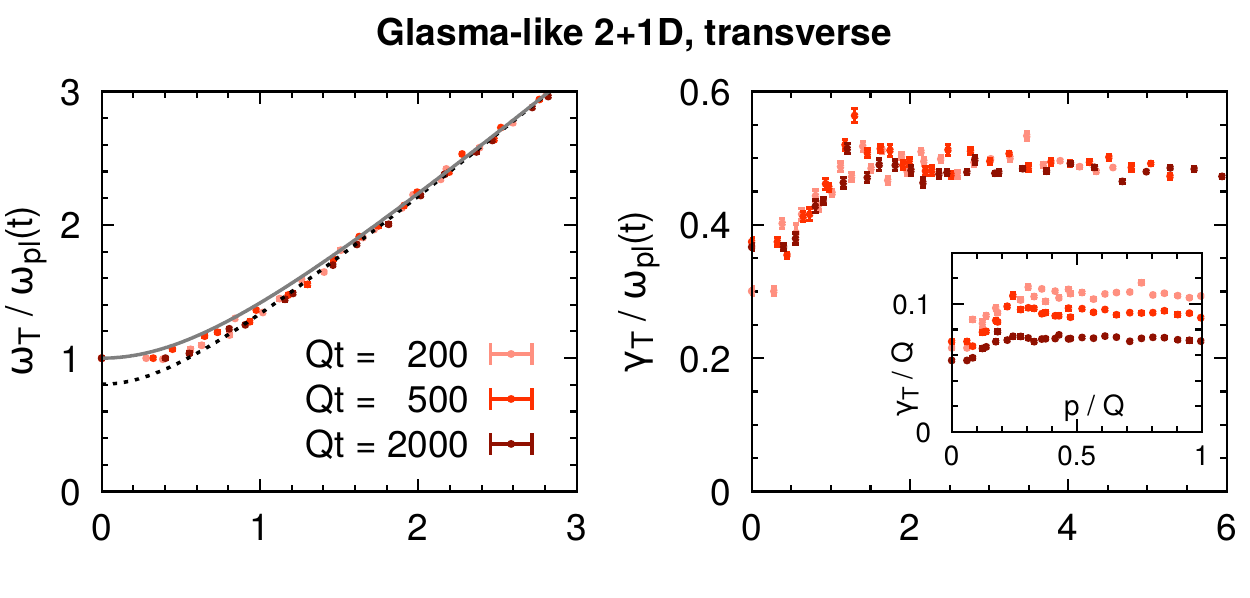}
	\includegraphics[scale=0.8]{\pToFigs/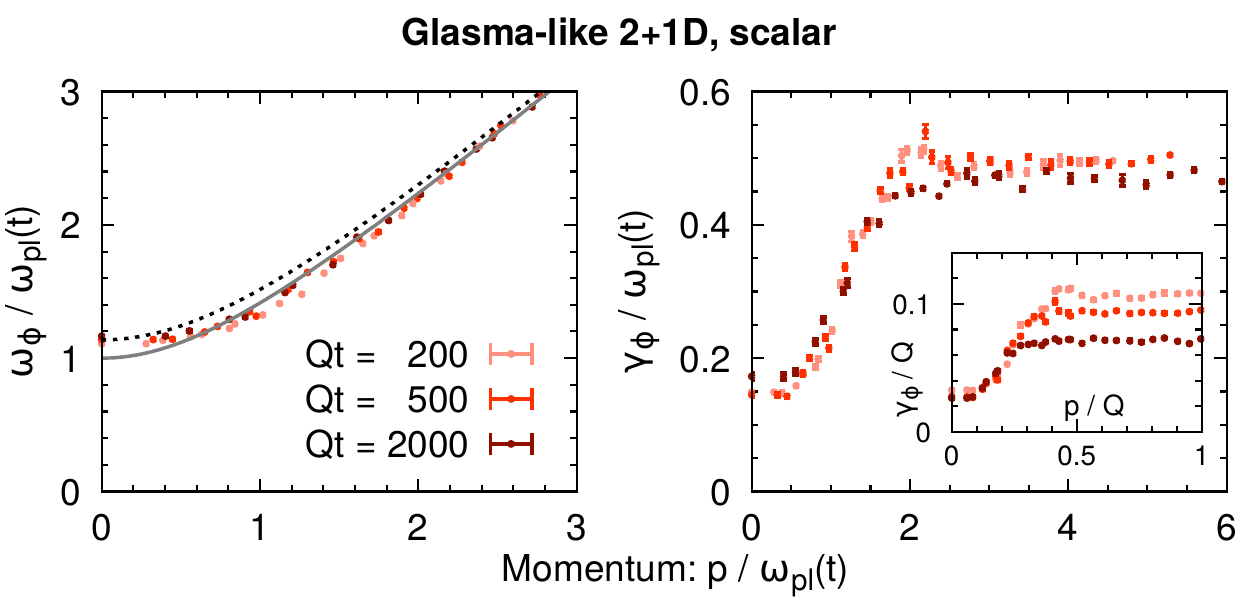}
	\caption{The peak position (dispersion relation) $\omega_\alpha(t,p)$ and the peak width (damping rate) $\gamma_\alpha(t,p)$ as functions of momentum $p$ at different times for ({\em top}:) the transversely polarized gluons of the usual $2+1$-dimensional theory and for ({\em center}:) the transversely polarized gluons and ({\em bottom}:) the scalar correlator of the Glasma-like $2+1$-dimensional theory.
	Dispersion relations, widths and momenta are normalized by the time dependent mass scale $\wplas(t)$ in the main plots and by the time independent energy scale $\Q$ in the insets.
	Additionally, we show, in the left panels, the same HTL dispersion relations $\omega_{T/\phi}^\HTL(p)$ and relativistic dispersions $\omega_{\rm{rel}}(p)$ as in \fig\ref{fig_2D_trans_ddF} as black dashed and continuous gray lines, respectively.
 }
	\label{fig_2D_w_g}
\end{figure}

\begin{figure}[t]
	\centering
	\includegraphics[scale=0.8]{\pToFigs/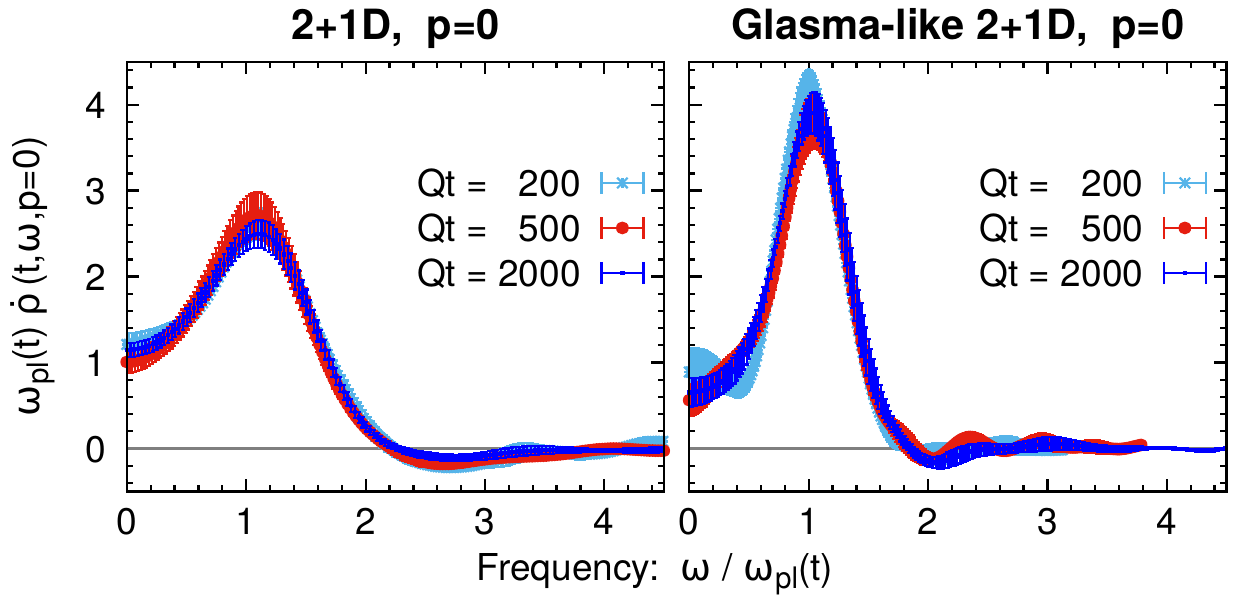}
	\caption{The dotted spectral function of gluonic excitations as a function of frequency for vanishing momentum $\wplas\dot{\rho}(t,\omega/\wplas,p{=}0)$ at different times, rescaled by the plasmon frequency such that they are made dimensionless. All curves fall on top of each other, showing the correlation functions follow a self-similar evolution with the only parameter $\wplas(t)$.
 }
	\label{fig_2D_w_diffT}
\end{figure}

\subsection{Time dependence of dispersion relations \& damping rates}
\label{sec:disp_gamma}

Now we will study how the dispersion relations $\omega_\alpha(t,p)$ and excitation widths $\gamma_\alpha(t,p)$ of the 2+1D systems depend on the time $t$. We recall that in the scaling solution (see \se\ref{sec:self-sim}) this can be used to obtain information about the dependence on the soft and hard scales, which follow separate power laws in $t$. 
Figure~\ref{fig_2D_w_g} shows our numerical results for the dispersion relations and damping rates at different times $Qt$ for the transverse gluon excitation in the 2+1D theory and for the transverse and scalar excitations of the Glasma-like 2+1D theory. They were extracted by fitting $h_{\text{Gauss}}(\omega, p)$ to the normalized transverse gluon and scalar excitations, as explained in \se\ref{sec:peaks}.
The main observation here is that all curves lie on top of each other when plotted in terms of the time dependent mass $\wplas(t)$. For the dispersion relation this is relatively trivial, and just shows that the functional form of the dispersion relation does not change in time and only depends on the current value of the mass. We have added HTL and relativistic dispersion relations as black dashed and continuous gray lines, respectively. We observe that both agree sufficiently well with the fitted values to the numerical data. Small deviations lie within the width of the excitation peaks, as discussed in \se\ref{sec:ColorPlots} and are visible in \fig\ref{fig_2D_trans_ddF}.

The $Qt$ depedence of the width, shown in the right panels of \fig\ref{fig_2D_w_g}, is much more interesting. Although the scaling is not quite as good as for the dispersion relation, the plots show that both the $p$-dependence and the magnitude of the damping rate follow the time dependence of the plasmon mass. This is different from the 3+1D case, where $\gamma(t)$ was found to decrease faster than $\wplas(t)$ with time \cite{Boguslavski:2018beu}. Thus, while for the 3+1D theory the quasiparticle peaks get narrower as the scale separation between the dynamical hard and soft scales grows with increasing time, this does not happen in the 2+1D theories: the width of the excitation peak relative to the mass remains constant.

As a consequence of this scaling,  correlators rescaled with appropriate powers of $\wplas(t)$ remain time-independent as functions of frequency $\omega/\wplas(t)$ and momentum $p/\wplas(t)$. This is demonstrated in \fig\ref{fig_2D_w_diffT} at the example of $\dot\rho_T$ at $p=0$ for both 2+1D systems. Over a wide range of times $\Q t = 200, 500, 2000$, the appropriately rescaled correlation function is indeed stationary. Note that we have not used self-similarity explicitly but merely rescaled all dimensionful quantities with appropriate powers of $\wplas(t)$. In fact, also in 2D classical thermal equilibrium the width of the excitations is of the order of the mass, as we show in Appendix~\ref{app_class}. This is suggestive of a general relation $\gamma_\alpha \sim \wplas$ for 2+1D systems. 

The behavior $\gamma_\alpha \sim \wplas$ should be contrasted with the expectation from HTL, where the width is expected to be proportional to the ``effective temperature of the soft modes''  $g^2 T_*(t)$. Indeed, in the 3+1D case we have observed~\cite{Boguslavski:2018beu} that the time dependence of $\gamma_\alpha$ is consistent with this expectation. For the 2+1D theories a simple parametric estimate would lead to the scaling $g^2 T_* \sim \Q(\Q t)^{-2/5}$ \cite{Boguslavski:2019fsb}, which is a much stronger time dependence than what we observe since $\wplas \sim \Q(\Q t)^{-1/5}$. Thus, the behavior of $\gamma_\alpha(t)$ in the 2+1D theories indicates that the phenomenon of plasmon decay happens in a qualitatively different way than in 3+1D. 

As a consequence, excitations at low momenta $p \lesssim m_D(t)$ for different times remain too short-lived to be considered as quasiparticles. Therefore, while an effective kinetic theory description may exist for larger momenta, its collision kernel, depending on the soft modes, must be determined nonperturbatively.


\section{Conclusion}
\label{sec:conclu}
In this paper we have studied the excitation spectrum of 2+1D classical gluodynamics in the self-similar regime. Our aim was to understand whether the boost invariant system created at the initial stages of ultrarelativistic heavy-ion collisions can be understood in a quasiparticle picture. We have studied two theories. The first one is a genuine 2+1D theory where the longitudinal direction is completely absent. The second theory is Glasma-like, in the sense that it is obtained from a 3+1D theory that is  invariant in the longitudinal direction. As a consequence, the gauge field in this direction becomes a scalar field.

One of our main observations is that excitations are broader in 2+1D than in isotropic 3+1D systems. This is seen in the damping rates, which correspond to the widths of the excitation peaks in statistical and spectral  correlation functions. 
In the 3+1D theory, the damping rate is a subleading effect, and an increasing scale separation between the hard and soft scales $\Lambda(t) \gg \wplas(t)$ (with increasing time $t$ in our case) leads to an increasing separation between plasmon mass and width, $\wplas(t) \gg \gamma(t)$. In the 2+1D cases, in contrast, the damping rate is of the order of the plasmon mass $\gamma(t) \sim \wplas(t)$ at all times.
This indicates that the damping rate is not determined by the hard degrees of freedom, but results from nonperturbative interactions between the soft gauge degrees of freedom. This is a qualitatively different dynamical picture than in three spatial dimensions and has the important consequence that gluonic transverse and longitudinal excitations below the mass scale are too short-lived to form quasiparticles.

Unlike the gauge fields, the scalar fields do have narrow excitations at low momenta. At larger momenta, the damping rates saturate and the transverse and scalar excitation peaks become identical.

Similarly to gluonic plasmas in 3+1D, we observe that the statistical and spectral correlation functions obey a generalized fluctuation-dissipation relation. However, in contrast to 3+1D, the excitation peaks have a non-Lorentzian shape, that is reasonably well described by Gaussian or hyperbolic secant distributions.

We have also performed simulations in classical thermal equilibrium in 2+1D and observed qualitatively similar behaviour as along the self-similar attractor. We interpret this as a sign that the qualitative features may be generic to 2+1D gauge theories. 

Our results indicate that an effective formulation of kinetic theory in 2+1 dimensions should indeed be possible for large momenta $p \gg \wplas$. However, this is complicated by the fact that in lower dimensions infrared effects become more important. 
In particular, in 2+1D the dynamically generated screening scale gets equal contribution from all momenta, whereas in 3+1D the dominant contribution comes from hard particles. As a consequence, the HTL approximation breaks down already at the Debye mass scale (instead of the magnetic scale in 3+1D). This picture is supported by our simulations. Therefore obtaining effective matrix elements for kinetic theory is a nonperturbative problem where the dynamics have to be deduced in a self-consistent way taking into account contributions from the plasmon mass scale. 

This work also contributes to the study of thermalization in heavy-ion collisions concerning the role of plasma instabilities, which is still under debate in the literature. 
In particular, there is a persistent discrepancy between classical Yang Mills simulations \cite{Berges:2013fga,Berges:2013eia} and corresponding HTL-theory calculations (including simulations and analytical calculations) \cite{Romatschke:2003ms,Rebhan:2004ur,Rebhan:2008uj,Attems:2012js}.
In the case of full 3+1D Yang-Mills theory, instabilities seem to play a much smaller role in the isotropization than in the case of HTL. This observation has been used as an argument to justify kinetic-theory descriptions without instabilities in phenomenological applications \cite{Kurkela:2015qoa,Kurkela:2018wud,Kurkela:2018vqr,Kurkela:2018xxd,Kurkela:2021ctp}. 
Our results show that, as expected, gluonic excitation modes at low momenta are not correctly described by HTL perturbation theory in 2+1D and 3+1D plasmas at extreme anisotropy. The presence of these nonperturbative corrections in the purely 2+1D case suggests that there may be significant nonperturbative corrections present also in systems with finite but large anisotropy, which are relevant to phenomenological applications and may alter the role of instabilities. For instance, this may cure some issues encountered in HTL perturbation theory where observables cannot be computed due to emerging instabilities in anisotropic systems.
To approach these questions in more detail, we plan to extend our simulations to expanding and anisotropic plasmas.


\begin{acknowledgments}
  We are grateful to A.\ Ipp, D.\ M\"uller, A.\ Rebhan and M.\ Strickland for valuable discussions.   J. P.\ acknowledges the support by the International Office of the TU Wien and would like to thank Institute for Theoretical physics for hospitality during part of this work.  This work has been supported by the European Research Council under grant no.~ERC-2015-CoG-681707,  by the EU Horizon 2020 research and innovation programme, STRONG-2020 project (grant agreement No 824093) and by the Academy  of  Finland,  project 321840.  This work was funded in part by the Knut and Alice Wallenberg foundation, contract number 2017.0036. The authors wish to acknowledge CSC - IT Center for Science, Finland, for computational resources. The computational results presented have been also achieved in part using the Vienna Scientific Cluster (VSC). The content of this article does not reflect the official opinion of the European Union and responsibility for the information and views expressed therein lies entirely with the authors.
\end{acknowledgments}


\appendix

\section{Formulas from the HTL framework}
\label{app_HTL2D}

\subsection{Polarization tensor}

In the HTL formalism, a crucial quantity is the polarization tensor $\Pi_{\mu\nu}$. For a non-Abelian $\SU(N)$ gauge theory in $d$ spatial dimensions, its leading order (LO) expression reads \cite{Braaten:1989mz,Blaizot:2001nr}
\begin{align}
 \label{eq_Pi_dDim}
 \Pi_{\mu\nu}(K) = g^2 \int_{\mbf p} V_\mu \frac{\partial \bar{f}_{\mbf p}}{\partial_{p_\beta}} \left(g_{\nu\beta} - \frac{V_\nu K_\beta}{K_\rho V^\rho + i\epsilon}\right),
\end{align}
where $\partial/\partial_{p_0} = 0$, with Minkowski metric $g_{\nu\beta} = \mrm{diag}(1,-\mbf 1)$, $\epsilon \rightarrow 0$, $V = (1,\mbf p/\omega_{\mbf p})$, $K = (\omega,\mbf k)$ and with the approximation $\omega_{\mbf p} \approx p$ for the dispersion relation, which is valid for sufficiently large momenta. In general, the frequency $\omega$ and momentum $\mbf k$ are independent. 
The polarization tensor is gauge invariant, symmetric $\Pi_{\mu\nu} = \Pi_{\nu\mu}$ and transverse with $K^\mu \Pi_{\mu\nu}(K) = 0$ for all $\nu$. For the latter to be true, the distribution function has to vanish at large momentum $\lim_{p_i \rightarrow \infty} f_{\mbf p} = 0$ for directions $i$, which is fulfilled in practice.

We have summed different bosonic contributions into the distribution function $\bar{f} = N_c \sum_{\lambda} f^{(\lambda)} =: N_c \dpol f(t,p)$, where each $f^{(\lambda)}$ corresponds to the distribution of the fields with non-longitudinal polarization $\lambda$. 
For the 2+1D theory, one has $\dpol = 1$ and the integration reads $\int_{\mbf p} = \int \ud^2 p/(2\pi)^2$. When considering the Glasma-like theory, the distribution function has the form $\fThrD(t,\mbf p) = 2\pi \delta(p_z)\, f(t,p_\perp)$, such that integration becomes identical to the 2+1D theory. Since the scalar contributions correspond to a polarization in $z$ direction, the distribution function $f(t,p)$ becomes an average over gauge and scalar distributions, i.e., $\dpol f(t,p) = f_G + f_\phi$ with $\dpol = 2$ in the Glasma-like theory. Note that the resulting expressions are the same for both 2+1D and Glasma-like theories with only a different $\dpol$.

Using the isotropy of $f(t,p)$ and performing an integration by parts, the expression \nr{eq_Pi_dDim} can be cast into the form
\begin{align}
 \label{eq_Pi_2D_mD}
 \Pi_{ij}(K) = m_D^2 \int_{0}^{2\pi}\frac{\ud \varphi}{2\pi} \,&\left[ \delta_{ij} - \frac{k_i v_j + k_j v_i}{-K_\rho V^\rho - i\epsilon} + \frac{(-\omega^2 + k^2) v_i v_j}{(-K_\rho V^\rho - i\epsilon)^2} \right],
\end{align}
with $i,j = 1, 2$ and 
\begin{align}
 \label{eq_DebyeM}
 m_D^2 = \dpol N_c \int \frac{\ud^2 p}{(2\pi)^2} \frac{g^2f(t,p)}{p}\,.
\end{align}
The transverse and longitudinal polarizations of $\Pi_{ij}(K)$ can be computed as
\begin{align}
 \Pi_T(\omega / k) = \frac{q^i\, \Pi_{ij}(K)\,q^j}{k^2}\,, \qquad \Pi_L(\omega / k) = \frac{k^i\, \Pi_{ij}(K)\,k^j}{\omega^2}\,,
\end{align}
with the transverse vector $\mbf q = (-k_2, k_1)$ such that $\mbf q \cdot \mbf k = 0$. 
Evaluating the residual integral \cite{Romatschke:2004jh}, one arrives at
\begin{align}
 \Pi_T(x) &= m_D^2\;x \left( x - \frac{x^2 - 1}{\sqrt{x+1} \sqrt{x-1}} \right) \\
 \Pi_L(x) &= m_D^2 \left( -1 + x\,\frac{x^2 - 1}{(x+1)^{3/2} (x-1)^{3/2}} \right)\,,
\end{align}
with $x = \omega / k$. 

For the Glasma-like theory, we also need the scalar component $\Pi_\phi = \Pi_{zz}$. To obtain it, we start from \eqref{eq_Pi_dDim} and perform an integration by parts
\begin{align}
 \Pi_\phi &= g^2 \int \frac{\ud^3 p}{(2\pi)^3}\, \bar{f}(t,\mbf p) \left[ \frac{\partial}{\partial_{p_z}} v_z + \frac{\partial}{\partial_{p_l}} \frac{v_z^2 k_l}{K_\rho V^\rho + i\epsilon}\right] \nonumber \\
 &= m_D^2 = const\,,
\end{align}
resulting from $p_z = k_z = 0$.

\subsection{Retarded propagator}

Next we consider the retarded propagator in temporal gauge $G^\HTL_{ij}(\omega, p)$, as employed in this work, and ask how these components of the polarization tensor emerge. Using a tensor decomposition as suggested in \res\cite{Kobes:1990dc,Romatschke:2003ms} with
\begin{align}
 A_{ij} = \delta_{ij} - \frac{p_i p_j}{p^2}\,, \quad B_{ij} = \frac{p_i p_j}{p^2}\,, \quad C_{ij} = \frac{n_i n_j}{n^2}\,,
\end{align}
where for the Glasma-like theory the normal vector is simply the unit vector in $z$ direction $\mbf n = \mbf e_z$, the propagator can be written as
\begin{align}
\label{eq:G_decomp}
 G^\HTL(\omega, p) = G_T^{\HTL} A + G_L^{\HTL}B + G_\phi^{\HTL}C,
\end{align}
with transverse, longitudinal and scalar polarizations
\begin{align}
 \label{eq_GR_HTL_T}
 G_T^{\HTL}(\omega, p) \,&= \frac{-1}{\omega^2 - p^2 - \Pi_T(\omega/p)} \\ 
 \label{eq_GR_HTL_L}
 G_L^{\HTL}(\omega, p) \,&= \frac{p^2}{\omega^2}\, \frac{-1}{p^2 - \Pi_L(\omega/p)} \\
 G_\phi^{\HTL}(\omega, p) \,&= \frac{-1}{\omega^2 - p^2 - m_D^2}\,.
 \label{eq_GR_HTL_N}
\end{align}
One immediately observes the same screening properties in the static limit as in the $3+1$-dimensional case, with
\begin{align}
 \lim_{\omega \rightarrow 0} \Pi_T = 0\,, \quad \lim_{\omega \rightarrow 0} -\Pi_L = m_D^2\,,
\end{align}
additional to the static screening of scalar fields with $\lim_{\omega \rightarrow 0} -\omega^2 + p^2 + m_D^2 = p^2 + m_D^2$. 
This gives the interpretation of $m_D^2$ as the Debye mass, as anticipated by its notation.

\subsection{Spectral function}
\label{app:spectral}

Each polarization of the spectral function can now be computed as the imaginary part of the respective retarded propagator%
\footnote{For the longitudinal spectral function, the correct $\epsilon$ prescription is to use $\Pi_L((\omega+i\epsilon)/p)$ in $G_L^{\HTL}$ while keeping the prefactor $p^2/\omega^2$ in \eq\nr{eq_GR_HTL_L} without this prescription. Otherwise one would obtain a term proportional to $\delta(\omega)$, which would lead to wrong sum rules.}
\begin{align}
 \rho_{\alpha}^\HTL(\omega,p) = 2\, \text{Im}\, G_{\alpha}^\HTL(\omega+i\epsilon,p)\,, 
\end{align}
with $\alpha = T, L, \phi$ denoting the polarization.
They satisfy the sum rules
\begin{align}
 \label{eq_sum_rules_2D_T}
 \dot{\rho}_{T}(t,\Delta t {=} 0,p) = \int_{-\infty}^{\infty} \dfrac{\ud \omega}{2\pi}\, \omega \rho_T(t,\omega,p) =&\, 1 \\
 \label{eq_sum_rules_2D_L}
 \dot{\rho}_{L}(t,\Delta t {=} 0,p) = \int_{-\infty}^{\infty} \dfrac{\ud \omega}{2\pi}\, \omega \rho_L(t,\omega,p) =&\, \dfrac{m_D^2}{p^2 + m_D^2} \\
 \label{eq_sum_rules_2D_phi}
 \dot{\rho}_{\phi}(t,\Delta t {=} 0,p) = \int_{-\infty}^{\infty} \dfrac{\ud \omega}{2\pi}\, \omega \rho_\phi(t,\omega,p) =&\, 1\,,
\end{align}
which are the same as in 3+1D for $T$ and $L$. 

The HTL spectral functions can be further split into a Landau damping part for $|\omega| < p$ and a quasiparticle part 
\begin{align}
\label{eq:rhoHTL_decomposition_app}
 \rho_{\alpha}^\HTL(\omega,p) = \rho_{\alpha}^{\mrm{Landau}}(\omega,p) + 2\pi Z_{\alpha}(p)\left[ \delta\!\left(\omega - \omega_{\alpha}^\HTL(p)\right) - \delta\!\left(\omega + \omega_{\alpha}^\HTL(p)\right) \right],
\end{align}
with the dispersion relations $\omega_{\alpha}^\HTL(p)$ and residues $Z_{\alpha}(p)$ of quasiparticle excitations as discussed below.

The Landau damping contributions have been written in the main text in \eqs\eqref{eq_Landau_damping_trans}-\eqref{eq_Landau_damping_phi}.
The dispersion relations of quasiparticle excitations can be deduced from \eqref{eq_GR_HTL_T}-\eqref{eq_GR_HTL_N} by looking for poles in the propagators for $\omega > p$. Defining $\pOvMD = p / m_D$, the dispersion relations can be calculated analytically
\begin{align}
\label{eq_HTL_trans_app}
 \omega_T^\HTL(p) \,&= m_D\,\sqrt{\frac{1 + 2\pOvMD^2 - 2 \pOvMD^4 + \sqrt{1 + 4\pOvMD^2}}{4 - 2\pOvMD^2}} \\ 
 \label{eq_HTL_long_app}
 \omega_L^\HTL(p) \,&= m_D\,\frac{1+\pOvMD^2}{\sqrt{2+\pOvMD^2}} \\
 \label{eq_HTL_scalar_app}
 \omega_\phi^\HTL(p) \,&= \sqrt{m_D^2 + p^2}\,,
\end{align}
where the dispersion relation for scalar particles is simply the relativistic dispersion of free particles. 
It is interesting to study their behavior at low and high momenta. These are
\begin{align}
 \label{eq_disprel_HTL_lowp}
 \omega_T^\HTL(p) \,&\overset{p\, \ll\, m_D}{\simeq} \sqrt{\wplas^2 + \frac{5}{4}\,p^2} \\
 \omega_L^\HTL(p) \,&\overset{p\, \ll\, m_D}{\simeq} \sqrt{\wplas^2 + \frac{3}{4}\,p^2} \\
 \label{eq_disprel_HTL_highp}
 \omega_T^\HTL(p) \,&\overset{p\, \gg\, m_D}{\simeq} \sqrt{p^2 + \masy^2} \\ 
 \omega_L^\HTL(p) \,&\overset{p\, \gg\, m_D}{\simeq} \sqrt{p^2 + \frac{\masy^4}{p^2}}\,,
\end{align}
with 
\begin{align}
 \wplas^2 = \frac{m_D^2}{2} \;, \qquad \masy^2 = m_D^2\,.
\end{align}
The leading asymptotic behavior is similar to the $d=3$ case. Like there, one has $\omega_T^\HTL(p) > \omega_L^\HTL(p)$ for $p>0$. The large-momentum behavior for transversely polarized quasiparticles agrees with the scalar dispersion, and hence, with a relativistic dispersion relation, which shows that $m_D$ can also be interpreted as the asymptotic mass. On the other hand, the longitudinally polarized quasiparticles do not have an asymptotic mass, just as in the 3+1-dimensional case. An important difference to the latter is, however, that their dispersion relation approaches the ultra-relativistic limit $\omega \simeq p$ considerably slower than in the $d=3$ case where the approach is exponential. 

The residues at the quasiparticle peaks are defined as
\begin{align}
 Z_{\alpha}(p) = -\left[ \frac{\partial G_{\alpha}^{-1}(\omega,p)}{\partial \omega} \right]^{-1}_{\omega = \omega_{\alpha}^\HTL(p)}
\end{align}
and read
\begin{align}
 Z_T(p) =&\, \left( 2\omega_T  - \frac{m_D^2}{p^2}\left( 2\omega_T - \frac{2\omega_T^2 - p^2}{\sqrt{\omega_T^2 - p^2}} \right) \right)^{-1} \\
 Z_L(p) =&\, \frac{\left( \omega_L^2 - p^2 \right)^{3/2}}{\omega_L^2\,m_D^2} \\
 Z_\phi(p) =&\, \frac{1}{2 \omega_\phi^2} \,,
\end{align}
with asymptotic values
\begin{align}
 Z_T(p) &\overset{p\, \ll\, m_D}{\simeq} \dfrac{1}{2\wplas} \\
 Z_L(p) &\overset{p\, \ll\, m_D}{\simeq} \dfrac{1}{2\wplas} \\
 Z_T(p) &\overset{p\, \gg\, m_D}{\simeq} \dfrac{1}{2p} \\
 Z_L(p) &\overset{p\, \gg\, m_D}{\simeq} \dfrac{m_D^4}{p^5}\,, \label{eq:ZL_pLarge}
\end{align}
where we dropped the $\HTL$ label and the dependence on momentum of the dispersion relations to shorten the expressions.
Thus, also in the 2+1D case one expects contributions from longitudinal quasiparticles to decrease at high momenta. However, it is a power law decrease, as opposed to the exponential decrease of 3+1D gauge theory. Note that the leading contribution of $Z_T$ for $p\, \ll\, m_D$ and $p\, \gg\, m_D$ and $Z_L$ for $p\, \ll\, m_D$ are the same in 2+1D and 3+1D. This implies that the sum rules \eqref{eq_sum_rules_2D_T}-\eqref{eq_sum_rules_2D_L} for the spectral functions are dominated by quasiparticle contributions for $p\, \ll\, m_D$ for both polarizations and for $p\, \gg\, m_D$ for transverse polarizations, while the longitudinal sum rule for $p\, \gg\, m_D$ is dominated by the Landau damping contribution.

\begin{figure}[t]
	\centering
	\includegraphics[scale=0.8]{\pToFigs/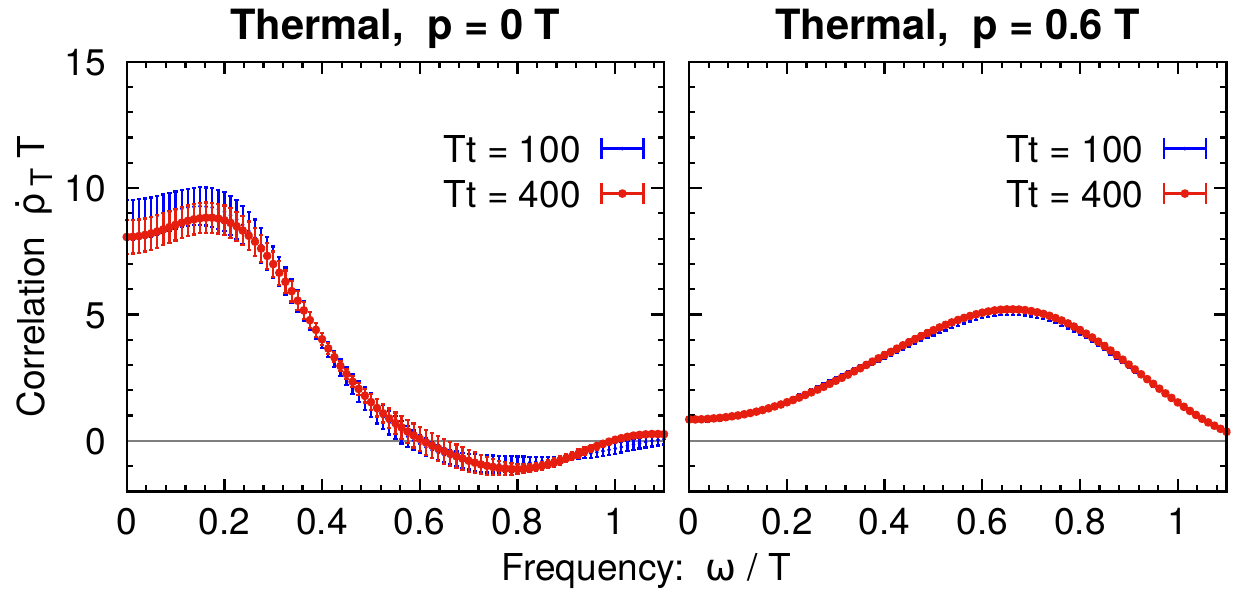}
	\caption{The correlator $\dot\rho_T$ as a function of frequency for momenta $p=0\,T$ (left) and $p=0.6\,T$ (right) for a 2+1D theory in classical thermal equilibrium. Time translation invariance is shown by comparing the correlations at two different times. One finds the same properties as for the non-equilibrium case considered in the main part of the paper. Note that in the right panel the curve for $T\,t=100$ is barely visible because it accurately coincides with the curve for $T\,t=400$.}
	\label{fig:2D_class}
\end{figure}

\section{Cross check: classical thermal equilibrium}
\label{app_class}

Here we compute the correlators of the 2+1D theory in frequency space in classical thermal equilibrium, and compare the results to the non-equilibrium systems discussed above. To obtain the classical thermal state, we initialize our system with $f(t=0, p) = T_0/p$ on a $256^2$ lattice with lattice spacing $T_0\,a=0.5$ and coupling $g^2=0.1\,T_0$. After the initialization we restore the Gauss law constraint and let the system evolve. We find that the equal-time correlators $\langle EE \rangle_{T/L}(t,\Delta t{=}0,p)$ become stationary for times $T_0\,t \gtrsim 20$, marking the onset of classical thermal equilibrium. Due to the small Debye mass $m_D^2 \sim g^2 T \ll T^2$, the resulting thermal state has almost the same temperature as the initial temperature parameter $T \approx T_0$.

Next, we extract correlation functions at different times $T\,t = 100$ and $400$ to demonstrate time-translation invariance explicitly, as should be the case in thermal equilibrium. Our results are shown for the transverse dotted spectral function $\dot{\rho}_T(t,\omega,p)$ in \fig\ref{fig:2D_class} for momenta $p=0$ (left) and $p=0.6\,T$ (right). One observes that the curves corresponding to different extraction times lie indeed on top of each other within statistical uncertainties. For higher momenta, where error bars are small, the curves can be barely distinguished by eye (right panel). This confirms that we have reached classical thermal equilibrium. 

Our numerical results qualitatively agree with our findings in the non-equilibrium systems above. We can summarize our most important findings as follows:
\begin{enumerate}
    \item We observe a fluctuation-dissipation relation, as should be trivially valid in thermal equilibrium.
    \item  The gluonic excitations are broad and their width is of the same order as the plasmon mass.
    \item The dotted spectral function is seen to approach a finite value for $\omega \to 0$ also in this thermal simulation. This indicates that this effect in the overoccupied system in Sec.~\ref{sec:smallomega} is not related to the violation of time reversal invariance, and should have some other physical origin.
\item The functional form is again non-Lorentzian and HTL does not provide a good description of the data (except for the longitudinal polarization at high momenta). 
\end{enumerate}
All of this indicates that the observed features in non-equilibrium 2+1D theories are not special to the considered self-similar attractor but also arise in the time-translation invariant setting of classical thermal equilibrium, and, presumably, for other non-equilibrium states.


\bibliographystyle{JHEP}
\bibliography{spires}

\end{document}